\documentclass[trackchanges,twocolumn]{aastex701}
\usepackage{amsmath,amssymb}
\usepackage{graphicx}
\usepackage{xcolor}
\usepackage{dcolumn}

\newcommand{\gm}{$g$-mode}
\newcommand{\be}{\begin{equation}}
\newcommand{\ee}{\end{equation}}
\newcommand{\beq}{\begin{eqnarray}}
\newcommand{\eeq}{\end{eqnarray}}

\newcommand{\gms}{$g$-modes}

\begin{document}

\title{Finite Temperature Effects on $g$-Modes of Warm Neutron Stars}

\author{David\,Morales-Zapien}
\email{David.morales-zapien02@student.csulb.edu} 
\affiliation{
Department of Physics \& Astronomy,
California State University Long Beach,
Long Beach, CA 90840, U.S.A.
}

\author{Prashanth\,Jaikumar}
\email{prashanth.jaikumar@csulb.edu}
\affiliation{
Department of Physics \& Astronomy,
California State University Long Beach,
Long Beach, CA 90840, U.S.A.
}

\author{Thomas\,Kl\"ahn}
\email{thomas.klaehn@csulb.edu}
\affiliation{
Department of Physics \& Astronomy,
California State University Long Beach,
Long Beach, CA 90840, U.S.A.
}

\begin{abstract}
We study the effect of temperature on composition-driven $g$-modes in the cores of warm neutron stars. Using a chiral $SU(2)_f$ sigma model, we construct isentropic equations of state and stellar configurations for hot and dense matter.  Weak-interaction rates are incorporated into a complex, frequency-dependent dynamical sound speed that approaches the frozen-composition limit for slow reactions and the beta-equilibrated limit for fast reactions. We solve self-consistently for the complex $g$-mode eigenfrequency and find that its real part depends non-monotonically on the nuclear symmetry-energy slope parameter $L$, reaching its largest values near $L\approx100$--$110~{\rm MeV}$. Increasing the entropy per baryon  accelerates chemical re-equilibration and suppresses the oscillation frequency. The associated damping rate reaches a maximum when the weak-interaction relaxation and mode timescales become comparable, signifying that dissipative effects also depend on $L$.
 Our results demonstrate that the thermal weak-interaction response and the density dependence of the symmetry energy jointly determine the $g$-mode spectrum of warm neutron stars.
\end{abstract}

\keywords{neutron stars}


\section{Introduction}\label{sec:level1}

The asteroseismology of neutron stars, encompassing studies of their oscillation mode spectrum and dynamical mode evolution, is intimately connected to the physics of dense matter and gravitational wave observations~\cite{Gittins_2025, Lozano_2022, J_2025,10.1093/mnras/stae1028,Ng_2021}. As is the case for massive stars and white dwarfs, neutron stars support a variety of oscillation modes. For neutron stars, some modes can become unstable to gravitational wave radiation, providing a fingerprint of the equation of state and composition of dense matter~\cite{AnderssonKokkotas1998, Andersson1998, PaschalidisStergioulas2017, GittinsAndersson2025}. The focus of this work is the $g$-mode, a non-radial fluid oscillation driven by buoyancy forces arising from temperature, entropy, and compositional gradients in the star~\cite{ReiseneggerGoldreich1992}. These modes may be excited to detectable amplitudes in ground-based interferometers~\cite{LVK2023} through strong tidal effects during the inspiral phase of neutron star mergers~\cite{YuWeinberg2017,KuanSuvorovKokkotas2021}, nonlinear hydrodynamics in proto-neutron stars formed in the aftermath of Type II supernovae~\cite{FerrariGualtieriPons2007,TorresForne2019}, or perturbations in the hot, dense post-merger remnant of a binary neutron star merger~\cite{BausweinJanka2012,ReyesBrown2018}.

The buoyancy-driven \gm\, is of particular interest as it provides a direct probe of the internal stratification of a neutron star. In contrast to pressure-driven modes, it is explicitly sensitive to local temperature, entropy, and compositional gradients. This sensitivity makes the \gm\, a useful diagnostic of the microphysical inputs governing the stellar interior. The microphysics enter through the difference between the equilibrium and adiabatic sound speeds, whose mismatch determines the Brunt–V\"ais\"al\"a frequency and thereby sets the conditions for global \gm\, support. Both core and crustal $g$-modes  have been extensively studied in the literature \cite{ReiseneggerGoldreich1992, Andersson2019, Lai1994, McDermott1988, Xia2025}, and are known to occupy distinct frequency bands with potentially observable signatures. Motivated by this, we focus on the temperature dependence of composition-driven \gm\, in warm neutron stars. With due attention to equilibration effects of weak reactions on the \gm\, we examine how its properties are controlled by the interplay between thermal gradients and composition gradients, as determined by key nuclear parameters, namely the density-dependent symmetry energy $S(n_B)$ and its slope $L(n_B)$, with $n_B$ the baryon density. These quantities, which compactly parameterize the energy per particle of $\beta$-equilibrated asymmetric $npe^-$ matter around saturation density, play a central role in neutron star structure and dynamics, influencing properties such as crustal physics, stellar radii, thermal relaxation, and dynamical response. Constraints from terrestrial experiments on their behavior can therefore be used to inform predictions for $g$-mode frequencies, as we demonstrate in this work. 

The role of finite temperature has been investigated in a variety of physical regimes and approximations. In newly born proto-neutron stars, during the first few seconds of evolution, large thermal gradients can significantly increase the $g$-mode frequency, while neutrino diffusion leads to rapid damping~\cite{2003MNRAS.342..629F}. During the subsequent cooling phase, when neutrino trapping remains important, calculations within the Cowling approximation have shown that both neutrino transport and relativistic effects can substantially modify the $g$-mode spectrum~\cite{2025EPJA...61..235T}. 
For convectively stable neutron stars, a different picture emerges. Under the assumption of a uniform temperature profile and slow reaction rates, composition-driven $g$-modes are no longer supported above $T{\sim}$10 MeV due to the lifting of degeneracy by thermal pressure contributions~\cite{Lozano_2022}. In contrast, $g$-modes associated with entropy gradients can persist, and may even be enhanced, at such temperatures~\cite{Gittins_2025}. Moreover, when finite reaction rates are taken into account, viscous effects can be incorporated through a dynamical sound speed, leading to an additional suppression of composition-driven $g$-modes already at  $T{\approx}$5 MeV~\cite{Zhao_2025}. Taken together, these results demonstrate that temperature can both enhance and suppress $g$-mode oscillations, depending on the underlying physical assumptions. This motivates further analysis of temperature effects on composition-driven $g$-modes.

For \gm\, studies, typical choices for the equation of state of dense nuclear matter have been either non-relativistic potential models~\cite{Semposki2021,Counsell2025} or relativistic mean field theories based on isoscalar and vector meson exchange~\cite{Tran2023,Guha2025,Zhao2022}. In this paper, we employ the chiral $SU(2)_f$ model of Sahu \& Ohnishi~\cite{sahu2000} that, while describing saturation properties of nuclear matter, also encodes the fundamental aspect of spontaneous breaking of chiral symmetry of QCD and its restoration at high density. In this way, the physics of chiral symmetry restoration affects the parameterization of nuclear matter, and consequently the value of the \gm\, frequency. We also incorporate finite temperature into this equation of state (EOS) via thermal occupation factors for the nucleons. 

The paper is structured as follows: Sec.~\ref{framework} describes our theoretical model of choice, the $SU(2)_f$ chiral sigma model, extended to finite temperatures, along with its parameter fits based on experimental constraints. In sec.~\ref{structure}, we compute the equilibrium structure of the star using the TOV equations and present the mass-radius curves of neutron star configurations. Secs.~\ref{gmodesection} and \ref{soundspeed} outline the calculation of \gms\,and sounds speeds, taking finite temperature into account. We discuss the \gm\, variation in regards to nuclear parameters in Sec.~\ref{symmetry}, gathering our conclusions in Sec.~\ref{conclusion}.

\section{Theoretical Framework}
\label{framework}
\subsection{Model Lagrangian}

Hadronic matter is modeled using the $SU(2)_f$ chiral sigma model 
where the strong interactions among nucleons, namely neutrons ($n$) and protons ($p$), are mediated 
by the exchange of scalar-isocalar ($\sigma$), vector-isoscalar ($\omega$), vector-isovector ($\rho$) and pseudoscalar-isovector($\pi$)
mesons. In terms of the chiral invariant $x^2$=$\sigma^2+\pi^2$ the interactions are described by the corresponding Lagrangian\cite{sahu1993,sahu2000, Sahu2004, Malik2017}:

\begin{equation}
\begin{aligned}
\mathcal{L} &=
\bar{\psi}_B 
\left[
i \gamma^\mu \partial_\mu
- g_\omega \gamma^\mu \omega_\mu - \frac{1}{2} g_\rho \, \vec{\rho}_\mu \cdot \vec{\tau} \, \gamma^\mu \right .
\\
&\quad
\left . - g_\sigma (\sigma + i \gamma_5 \vec{\tau} \cdot \vec{\pi})
\right ] \psi_B
+ \frac{1}{2} (\partial_\mu \vec{\pi} \cdot \partial^\mu \vec{\pi}
+ \partial_\mu \sigma \, \partial^\mu \sigma)
\\
&\quad
- \frac{1}{4} F_{\mu\nu} F^{\mu\nu}
- \frac{1}{4} \vec{R}_{\mu\nu} \cdot \vec{R}^{\mu\nu}
\\
&\quad
- \frac{\lambda}{4} (x^2 - x_0^2)^2
- \frac{\lambda b}{6 m^2} (x^2 - x_0^2)^3
- \frac{\lambda c}{8 m^4} (x^2 - x_0^2)^4
\\
&\quad
+ \frac{1}{2} g_\omega^2 x^2 (\omega_\mu \omega^\mu)
+ \frac{1}{2} m_\rho'^2 \, \vec{\rho}_\mu \cdot \vec{\rho}^\mu
+\eta \frac{1}{2} g_\rho^2 x^2 \, \vec{\rho}_\mu \cdot \vec{\rho}^\mu\,,
\end{aligned}
\end{equation}

where $\psi_B$ denotes the $SU(2)$ nucleon isospin doublet, and $g_\sigma$, $g_\omega$, and $g_\rho$ are the coupling constants that characterize the interaction strengths between the nucleons and the corresponding meson fields. The constant $c$ and $b$ are couplings for the self interaction of the scalar fields.

Spontaneous chiral symmetry breaking ($\chi$SB) gives rise to a dynamical mass for the $\omega$ meson through its interactions with the $\sigma$ and $\pi$ fields. In this framework, the vacuum expectation value $x_0$ of the $\sigma$ field determines the nucleon mass $(m)$ and the meson masses $(m_\sigma, m_\omega)$.
\begin{equation}
    m = g_\sigma x_0\,, \quad m_\sigma = \sqrt{2\lambda}x_0\,, \quad m_\omega = g_\omega x_0
\end{equation}
where $\lambda = (m_\sigma^2 - m_\pi^2)/(2f_\pi^2)$ and $f_\pi$ is the pion decay constant. In this work we will not consider the possibility of pion condensation, although the model may support it for a different choice of parameters. 
Furthermore, the presence of a $\sigma$–$\rho$ coupling introduces an additional contribution to the $\rho$-meson mass arising from $\chi SB$, leading to

\begin{equation}
    m_\rho ^2= {m'}_\rho ^2 + \eta g_\rho^2x^2_0 \,.
\end{equation}

\subsection{Equation of State}
We employ the relativistic mean-field approximation in which the spatial and temporal variations of the meson fields are neglected and the fields take on the value of their ground state expectation value. This results in nonzero values for only $\sigma, \omega_0$ and  $\rho_0$. The finite temperature energy density $\epsilon$ and pressure $p$ for neutron star matter is then given by:

\begin{equation}
\begin{aligned}
\epsilon &=
 \frac{m^2}{8 c_\sigma} (1 - y^2)^2
- \frac{b}{12 c_\sigma c_\omega} (1 - y^2)^3 
\\
&\quad
+ \frac{c}{16 m^2 c_\sigma c_\omega^2} (1 - y^2)^4
+ \frac{1}{2} m_\omega^2 \, \omega_0^2 \, y^2
\\
&\quad
+ \frac{1}{2} m_\rho^2 \left[ 1 - \eta (1 - y^2) \frac{c_\rho}{c_\omega}  \right] (\rho_0^3)^2
\\
&\quad
 +\frac{1}{\pi^2} \sum_{i=n,p,e} \int_0^{\infty}  dk\,k^2 \,  \sqrt{k^2 + m^{*2}} \left[f_i +\bar{f_i} \right ]
\end{aligned}
\end{equation}

\begin{equation}
\begin{aligned}
p &=
 -\frac{m^2}{8 c_\sigma} (1 - y^2)^2
+ \frac{b}{12 c_\sigma c_\omega} (1 - y^2)^3 
\\
&\quad
- \frac{c}{16 m^2 c_\sigma c_\omega^2} (1 - y^2)^4
+ \frac{1}{2} m_\omega^2 \, \omega_0^2 \, y^2
\\
&\quad
+ \frac{1}{2} m_\rho^2 \left[ 1 - \eta (1 - y^2) \frac{c_\rho}{c_\omega}  \right] (\rho_0^3)^2
\\
&\quad
 +\frac{1}{3\pi^2} \sum_{i=n,p,e} \int_0^{\infty}  dk\,\frac{k^4}{\sqrt{k^2 + m^{*2}}} \left[f_i +\bar{f_i} \right ]
\end{aligned}
\end{equation}

where $y = x/x_0$ and $f_i$ and $\bar{f_i}$ are the fermi dirac distribution of particle $i$. In the case of the nucleons these are given by:

\begin{equation}
    \begin{aligned}
    f_i &= \frac{1}{\exp\left(\frac{\sqrt{k^2 + {m^*}^2} - \nu}{ T}\right) + 1}\\
    \bar{f_i} &= \frac{1}{\exp\left(\frac{\sqrt{k^2 + {m^*}^2} + \nu}{ T}\right) + 1}
\end{aligned}
\end{equation}
and $\nu $ is the effective chemical potential of each baryon:
\begin{equation}
    \nu_i = \mu_i - g_\omega\omega - I_{3\rho}g_\rho \rho_0^3
\end{equation}

Where $I_{3\rho}$ is the isospin of the proton (1/2) or the neutron (-1/2). For electrons the kinetic terms are given by the substitution $m^* \rightarrow m_e$ and $\nu_i \rightarrow \mu_e$. We also define the coupling constants $c_i = g_i^2/m_i^2. $The values of the meson mean fields are obtained by solving the self-consistent mean field equations, derived from the requirement that the energy density be stationary with respect to variations in each field. The field equations are therefore given by:

\begin{equation}
\label{sigma_equation}
\begin{split}
    &(1 - y^2) -  \frac{b}{m^2 c_\omega}{} (1 - y^2)^2 + \frac{c}{m^4 c_\omega^2} (1 - y^2)^3 \\&+\frac{2c_{\sigma}c_{\omega}n_B^2}{m^2y^4}+ \frac{2\eta c_\sigma c_\rho m_\rho^2 ({\rho^0_3})^2}{c_\omega m^2}\\&
- \frac{c_\sigma \gamma}{\pi^2} \sum_{i=n,p} \int_0^\infty \frac{k^2 \, dk \, (f_i(T) + \bar{f}_i(T))}{\sqrt{k^2 + m^{*2}}} 
= 0,
\end{split}
\end{equation}

\begin{equation}
\label{omega_eqn}
    \omega_0 = \frac{c_\omega n_B}{g_\omega y^2}
\end{equation}

\begin{equation}
\label{rho_equation}
    m_\rho^2 \left [ 1- \eta (1- y^2) \frac{c_\rho}{c_\omega}\right] \rho_3^0 = \frac{1}{2}g_\rho(n_p -n_n)
\end{equation}

where the individual particle densities are given by 
\begin{equation}
n_i = \frac{\gamma}{(2\pi)^3} \int_0^{\infty} d^3k \, [f_i(T) - \bar{f}_i(T)],
\end{equation}
For background neutron star matter, we must also impose the constraints of charge neutrality, $\beta$-equilibrium and total baryon density:
\begin{equation}
\label{charge}
    n_e = n_p \,,\quad \mu_n = \mu_p + \mu _e\, \,,\quad n_B = n_p + n_n
\end{equation}

\noindent Numerical simulations of protoneutron star evolution indicate that within approximately 100~ms after core bounce, convective processes establish an almost uniform entropy per baryon, $s$, throughout the stellar interior~\cite{Pascal2022}. At this stage, for the protoneutron star, which contains trapped neutrinos, one may set $s = 1$ where we will use the units of $k_B = 1$. On a timescale of 10--20~s, the neutrinos diffuse from the star but leave behind much of their energy, causing the core temperature to rise, leading to a subsequent stage with $s = 2$~\cite{Kunkel_2025, Prakash_1997, Steiner2000}. Finally, the protoneutron star begins to cool through the thermal emission of neutrinos, eventually resulting in a cold neutron star described by $s = 0$ \cite{Mariani_2017}. As a finite-temperature study of g-modes may have implications for protoneutron stars, we  will explore the region where $s \leq 2$, and the entropy per baryon $s$ is given as in~\cite{Shen_1998} by

\begin{equation}
\label{entropy}
\begin{split}  
s = &\sum_i \frac{\gamma}{2\pi^2n_B} \int_0^\infty dk\, k^2 \left[
    - f_i(k, T) \ln f_i(k, T) \right .
    \\&
    - \left . \left(1 - f_i(k, T)\right) \ln\left(1 - f_i(k, T)\right) - \bar{f}_i(k, T) \ln \bar{f}_i(k, T)\right.\\
    &\left. 
    - \left(1 - \bar{f}_i(k, T)\right) \ln\left(1 - \bar{f}_i(k, T)\right)
\right].
\end{split}
\end{equation}

The field equations (\ref{sigma_equation}), \ref{omega_eqn} (\ref{rho_equation}) along with the 4 constraint conditions of \ref{charge} and \ref{entropy} can thus be solved self consistently for the unknowns $y, \omega_0, \rho_3^0 , \mu_n, \mu_p,\mu_e,T$ as a function of density, allowing us to solve for all relevant thermodynamic quantites.
\subsection{Parameter Sets}

The model contains six free parameters: $c_\sigma$, $c_\omega$, $c_\rho$, $b$, $c$, $\eta$. These parameters are determined by fitting to the empirical properties of symmetric nuclear matter (SNM) at saturation. In particular the parameters $c_{\sigma}$, $c_\omega$, $b$, $c$ are fit to reproduce the values of the nucleon effective mass $m^* $, the binding energy per nucleon $E_0 $, and the nuclear incompressibility $K$ at saturation density $n_0 = 0.153~\text{fm}^{-3}$ for symmetric nuclear matter at zero temperature. In addition, $c_\omega$ is constrained by the relation $f_\pi = 1/\sqrt{c_\omega}$, which ties the vector coupling directly to the pion decay constant. 
\par Once these SNM parameters are determined, the remaining parameters $c_\rho$ and $\eta$ control the behaviour of isospin-asymmetric matter. The energy per baryon $E = \epsilon / n_B$ may be expanded around symmetric matter in powers of the isospin asymmetry $t = (n_n - n_p)/(n_n +n_p)$ as:

\begin{equation}
    E(n_B, t) = E_0(n_B, 0) + S_N(n_B)t^2 + \ldots,
\end{equation}
where $E_0$ is the binding energy per nucleon for symmetric nuclear matter and $S(n_B)$ represents the symmetry energy. 

The symmetry energy can itself be expanded as a Taylor series about the saturation density $n_0$:
\begin{equation}
    S(n_B) = S(n_0)
    + \left(\frac{dS}{dn_B}\right)_{n_0}(n_B - n_0)
    + ...
\end{equation}
with the symmetry energy slope $L$ at saturation
\begin{equation}
\label{sym_energy}
    L = 3n_0\left(\frac{dS}{dn_B}\right)_{n_0}
\end{equation}
Within the framework of our model, the symmetry energy takes the form
\begin{equation}
    S(n_B) = \frac{k_F^2}{6\sqrt{k_F^2 + {m^*}^2}} 
    + \frac{c_\rho k_F^3}{12\pi^2} 
    \left( \frac{m_\rho}{m^*_\rho} \right)^2,
\end{equation}
where the Fermi momentum and the effective $\rho$-meson mass are given by
\begin{equation}
    k_F = \left( \frac{3\pi^2 n_B}{2} \right)^{1/3},
    \qquad
    {m_\rho^*}^2 = m_\rho^2 
    \left[ 1 - \eta (1 - y^2) 
    \left( \frac{c_\rho}{c_\omega} \right) \right].
\end{equation}
whereas the symmetry energy slope $L$ is given by: 
\begin{equation}
\begin{split}
L &= 
\left[\frac{k_F^2}{3E_k} - \frac{k_F^4}{6E_k^3}\right]
- n_0\,\frac{k_F^2 m^2 y}{2E_k^3}\,\frac{\partial y}{\partial n_B}
+ \frac{c_\rho k_F^3}{4\pi^2}\!\left(\frac{m_\rho^*}{m_\rho}\right)^{-2}\\
&
- \frac{n_0 c_\rho^2 k_F^3 \eta y}{2\pi^2 C_\omega } \left (\frac{m_\rho^*}{m_\rho}\right )^{-4}
\,\frac{d y}{d n_B}\\[3pt]
\end{split}
\end{equation}
We adopt a nuclear matter incompressibility of $K_0 = 250~\mathrm{MeV}$, consistent with the value $K_0 = 240 \pm 20~\mathrm{MeV}$ extracted from analyses of the isoscalar giant monopole and isoscalar giant dipole resonances in finite nuclei~\cite{Garg_2018}. For the symmetry energy we take $S_0 = 32.0~\mathrm{MeV}$, a value supported by terrestrial experiments~\cite{Li_2013, Oertel_2017, nuclear_sym} and consistent with chiral effective field theory predictions, $S_0 = 31.7 \pm 1.1~\mathrm{MeV}$~\cite{Drischler_2020}.

While $K_0$ and $S_0$ are relatively well constrained, the density 
dependence of the symmetry energy, which is encoded in the slope parameter $L$, remains highly uncertain. Recent precision measurements of the neutron skin thickness have produced mutually inconsistent 
constraints. As summarized in a recent review~\cite{Tagami2022}, the 
CREX result for $^{48}\mathrm{Ca}$ favors a relatively soft symmetry energy,
\begin{equation}
    0 \lesssim L \lesssim 51~\mathrm{MeV},
\end{equation}
whereas the PREX-II measurement on $^{208}\mathrm{Pb}$ suggests a much stiffer symmetry energy,
\begin{equation}
    76~\mathrm{MeV} \lesssim L \lesssim 165~\mathrm{MeV}.
\end{equation}
These two experimental bands do not overlap, underscoring the unresolved tension in neutron-skin measurements and their mapping to symmetry-energy parameters. However, recent analyses have argued that the apparent discrepancy may be overstated. In particular, \citet{Lattimer2023} showed that, once correlations between the neutron skin, symmetry energy, and model systematics are fully accounted for, the CREX and PREX-II results are not statistically incompatible and can be simultaneously accommodated within a consistent theoretical framework.

Astrophysical observations provide an important, complementary constraint. A recent meta-analysis of 24 independent studies of neutron-star masses, radii, and tidal deformabilities finds~\cite{Li2021}
\begin{equation}
    L \approx 57.7 \pm 19~\mathrm{MeV} 
    \quad ,
\end{equation}
a range that lies between the CREX and PREX-II determinations and is consistent with many microscopic nuclear-theory calculations. As current terrestrial and astrophysical measurements span a wide range of possible values of $L$, it is useful to investigate this full parameter space. In this work, we therefore examine how varying $L$ impacts the resulting equation of state and neutron-star properties. The inclusion of the $\sigma\!-\!\rho$ coupling provides the flexibility to adjust $L$ while keeping the underlying parameter set fixed. However, for any chosen parameter set, the model imposes a strict lower bound on the symmetry-energy slope $L$. As we show below, this limit arises from the requirement that the effective $\rho$-meson mass remain real and positive, ensuring physical solutions for the symmetry energy. To obtain this limit, we note that the expression for the interaction part of the symmetry energy is
\begin{equation}
S - S_{\text{kin}} =
\frac{C_\rho k_F^3}{12\pi^2}
\frac{1}{1 - \eta(1 - y^2)\,C_\rho/C_\omega}\,.
\end{equation}

To simplify notation, we define
\begin{equation}
\beta \equiv \frac{12\pi^2}{k_F^3}\,(S - S_{\text{kin}}),
\qquad
\alpha \equiv \frac{\eta(1 - y^2)}{C_\omega},
\end{equation}
which allows the preceding equation to be written compactly as
\begin{equation}
\beta = \frac{c_\rho}{1 - \alpha c_\rho}\,.
\end{equation}
from which it follows that
\begin{equation}
C_\rho = \frac{\beta}{1 + \beta\alpha}
       = \frac{\beta}{1 + \beta\,\eta(1 - y^2)/C_\omega}.
\end{equation}

This closed-form relation shows that \(c_\rho\) diverges when the denominator
vanishes, i.e.
\begin{equation}
1 + \beta\alpha = 0
\quad\Rightarrow\quad
\eta = \eta_{\text{crit}} \equiv
-\frac{c_\omega}{\beta(1 - y^2)}.
\end{equation}
At this critical coupling \(\eta_{\text{crit}}\), the effective
$\rho$-meson mass tends to zero and $C_\rho \to \infty$.
This condition therefore defines the boundary of the physically allowed region
in parameter space: for $\eta < \eta_{\text{crit}}$, the squared effective mass
$m_\rho^{*2}$ becomes negative, leading to unphysical solutions.

The same critical point sets the minimum physically allowed value of \(L\).
From the definition of the slope parameter,
\begin{equation}
\begin{split}
L &= 
L_{\text{kin}}
+ \frac{c_\rho k_F^3}{4\pi^2}\!\left(\frac{m_\rho^*}{m_\rho}\right)^{-2}
- \frac{n_0 c_\rho^2 k_F^3 \eta y}{2\pi^2 c_\omega } \left (\frac{m_\rho^*}{m_\rho}\right )^{-4}
\,\frac{d y}{d n_B} \\[3pt]
&= L_{\text{kin}} + 3(S_0 - S_{\text{kin}}) - D\,\beta^2\,\eta,
\end{split}
\end{equation}
the lower bound follows directly by substituting $\eta = \eta_{\text{crit}}$:
\begin{equation}
L_{\text{min}}
= L_{\text{kin}}
+ 3(S_0 - S_{\text{kin}})
- D\,\beta^2\,\eta_{\text{crit}}.
\end{equation}

At this point, $m_\rho^*\!\to\!0$ and any further decrease in \(L\) drives the system into an unphysical regime. Thus, \(L_{\text{min}}\) represents the absolute lower limit permitted by the internal consistency of the mean-field framework. For the parameter set we employ this limit occurs ar $L_{min} \approx  49$ MeV. Example of coupling constants for $L = 50, 100, 150$ MeV are shown in Table ~\ref{tab:Lsnm-params}.

\begin{figure}[t]
\includegraphics[width=\columnwidth]{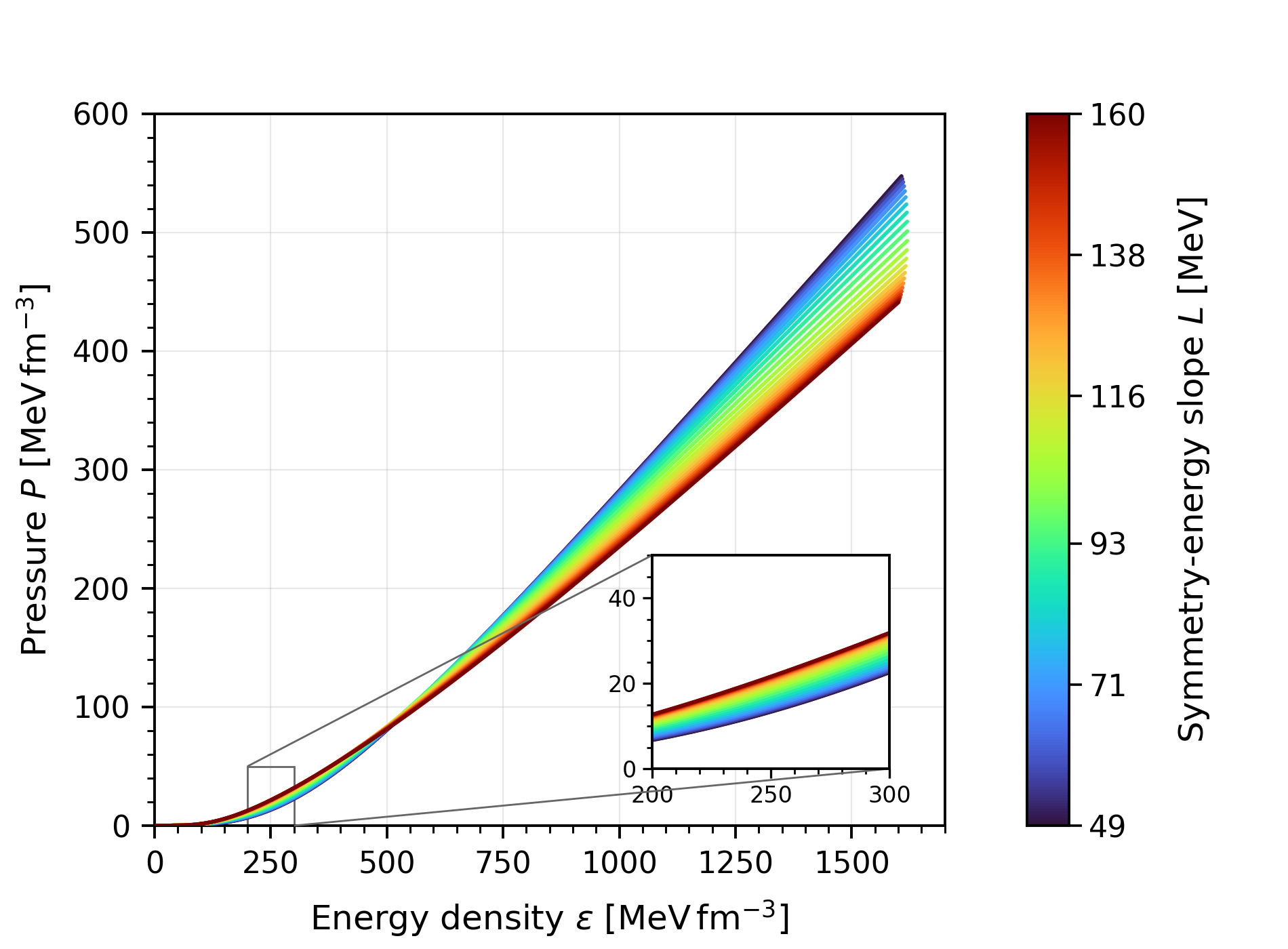}
\caption{\label{fig:eoscolor}
Equation of state for a representative value of $s = 0.1$ for several values of the symmetry energy slope $L$. The curves exhibit a crossing behavior: at low densities, larger $L$ corresponds to a stiffer equation of state, while at higher densities the trend reverses, leading to a softer equation of state.}
\end{figure}

\begin{figure}[t]
\includegraphics[width=\columnwidth]{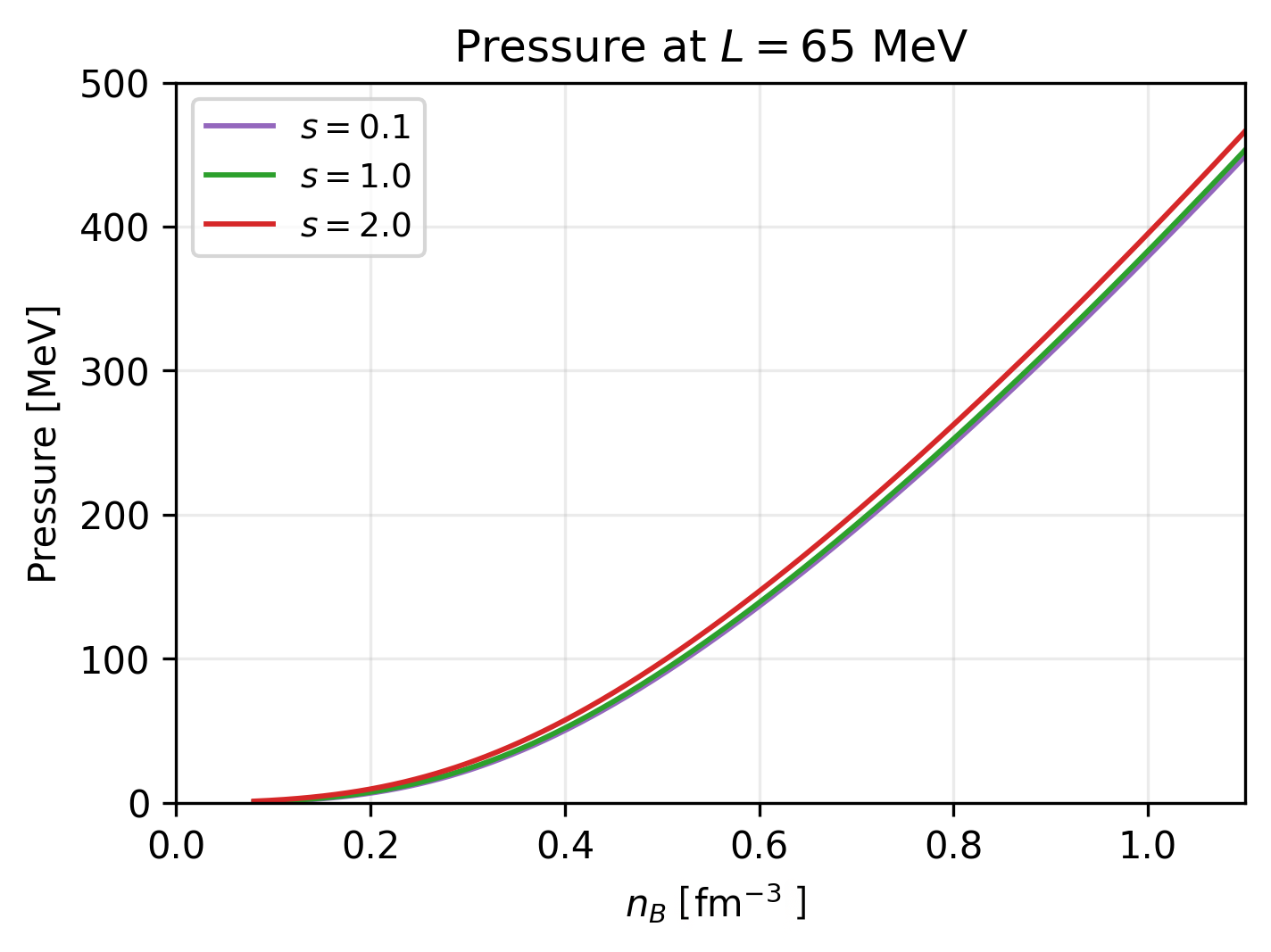}
\caption{\label{fig:pressures}
Pressure as a function of baryon density $n_B$ for several values of the entropy per baryon $s$. Finite-temperature effects provide additional thermal pressure, shifting the equation of state to higher pressures at fixed density, with the largest deviations occurring at higher densities.
}
\end{figure}
The equation of state obtained from the $SU(2)$ chiral model, extended to finite temperature, is shown in Fig.~\ref{fig:eoscolor} for fixed $s$=0.1 for different values of $L$. Near saturation density, smaller values of $L$ produce a softer EOS. However, the curves intersect at roughly $3.5\,\rho_0$, beyond which the EOS corresponding to low $L$ becomes significantly stiffer. This high-density stiffening arises because small values of $L$ require a much larger $\rho$-meson coupling $c_\rho$. The enhanced value of $c_\rho$ amplifies the kinetic $\rho$-meson contribution to the energy density and pressure, which in turn drives the EOS to become stiffer at high density.
\noindent

\newcolumntype{d}[1]{D{.}{.}{#1}} 
\begin{table}[t]
  \caption{Model parameters fixed from SNM properties:
  $E_0=-16$ MeV, $K=250$ MeV, $y=m^\ast/m=0.864$ at $n_0=0.153~\mathrm{fm}^{-3}$
  $B=b/m^2$ and $C=c/m^4$.}
  \label{tab:snm-params}
  \begin{ruledtabular}
  \begin{tabular}{d{1.3} d{1.3} d{2.3} d{1.3}}
    \multicolumn{1}{c}{$C_\sigma$} & \multicolumn{1}{c}{$C_\omega$} & \multicolumn{1}{c}{$B$} & \multicolumn{1}{c}{$C$} \\
    \multicolumn{1}{c}{(fm$^{2}$)} & \multicolumn{1}{c}{[fm$^{2}$]} & \multicolumn{1}{c}{[fm$^{2}$]} & \multicolumn{1}{c}{[fm$^{4}$]} \\
    \hline
     7.011 & 1.767 & -5.616 & 0.404 \\
  \end{tabular}
  \end{ruledtabular}
\end{table}

\begin{table}[t]
\caption{Values of the isovector coupling constants $c_\rho$ and $\eta$ corresponding to the representative symmetry-energy slopes $L=50$, $100$, and $150~\mathrm{MeV}$. The symmetric nuclear matter parameters are fixed as in Table~1, and the symmetry energy at saturation is held fixed at $S_0=32~\mathrm{MeV}$.}
  \label{tab:Lsnm-params}
  \begin{ruledtabular}
  \begin{tabular}{d{1.3} d{1.3} d{2.3} }
    \multicolumn{1}{c}{$L$} & \multicolumn{1}{c}{$C_\rho$} & \multicolumn{1}{c}{$\eta$}  \\
    \multicolumn{1}{c}{[MeV]} & \multicolumn{1}{c}{[fm$^{2}$]} & \multicolumn{1}{c}{[fm$^{2}$]}  \\
    \hline
     50 & 135.858 & -1.342  \\
     100 & 3.597 & 0.542  \\
     150 & 1.823 & 2.426  \\
  \end{tabular}
  \end{ruledtabular}
\end{table}


\section{EQUILIBRIUM STRUCTURE}
\label{structure}
By treating the star as a perfect fluid, one obtains the equations of hydrostatic equilibrium in general relativity, commonly known as the Tolman–Oppenheimer–Volkoff (TOV) equations:
\begin{equation}
\frac{dp(r)}{dr}
=
-\frac{\left[\varepsilon(r)+p(r)\right]
\left[m(r)+4\pi r^{3}p(r)\right]}
{r^{2}\left[1-\dfrac{2m(r)}{r}\right]},
\label{eq:TOV_pressure}
\end{equation}

\begin{equation}
\frac{dm(r)}{dr}
=
4\pi r^{2}\varepsilon(r),
\label{eq:TOV_mass}
\end{equation}

\begin{equation}
\frac{d\nu(r)}{dr}
=
\frac{2\left[m(r)+4\pi r^{3}p(r)\right]}
{r\left[1-\dfrac{2m(r)}{r}\right]},
\label{eq:TOV_metric}
\end{equation}
These equations are simultaneously integrated with the boundary conditions $m(r=0) = 0 $ , $\epsilon(r=0) = \epsilon_c$ at the center and $p(r=R) = 0$ at the surface. In the Schwarzschild metric, the function $\lambda(r)$ is given by $e^{\lambda(r)} = \frac{r}{r-2m(r)}$ and $e^{\nu(r)}$ is normalized such that at the surface it is equal to the exterior Schwarzchild  metric, $e^{\nu(R)} = (1- 2M/R)$. 
The total baryon number is conserved during the quasi-static evolution of proto-neutron stars \cite{Raduta_2020}, making it convenient to evolve the enclosed baryon number $a(r)$ alongside the TOV equations. The baryon number satisfies
\begin{equation}
\frac{da(r)}{dr}
=
4\pi r^{2} n_B(r)
\left(1-\frac{2m(r)}{r}\right)^{-1/2},
\label{eq:baryon_number}
\end{equation}
which can be solved simultaneously with Eqs.~\eqref{eq:TOV_pressure}, \eqref{eq:TOV_mass}, and \eqref{eq:TOV_metric}.
The baryonic mass of the star is then defined as
\begin{equation}
M_B = a(R)\,m_n,
\end{equation}
where $m_n$ is the nucleon mass, and $M_B$ generally differs from the gravitational mass $m(R)$ due to the star’s gravitational binding energy.

\begin{figure}[h]
\includegraphics[width=\columnwidth]{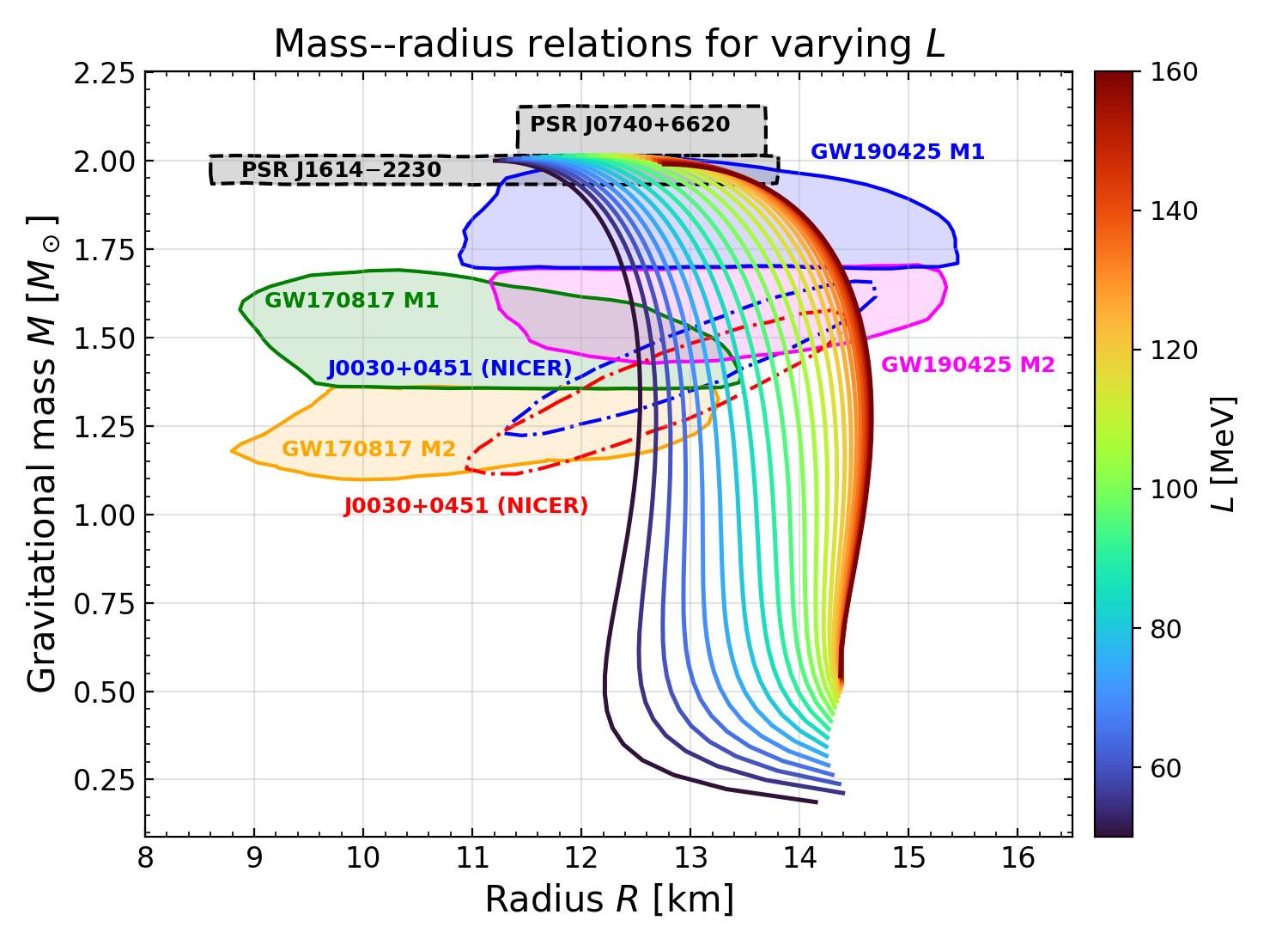}
\caption{\label{fig:MR_L}
Mass--radius relations for several values of the symmetry-energy slope parameter $L$.
The theoretical curves are compared with observational constraints from the massive pulsars PSR J0740+6620 \cite{Fonseca_2021,Riley_2021} and PSR J1614--2230 \cite{Demorest2010,MajidSharif2020}, as well as NICER measurements of PSR J0030+0451 \cite{Riley_2019,Miller_2019}.
Also shown are the 90\% confidence intervals inferred from gravitational-wave observations of binary neutron star mergers \cite{Abbott_2017,Abbott_2020,Abbott_2018}. The model is consistent with constraints for L $\lesssim$ 80 MeV.}
\end{figure}

\begin{figure}[t]
d\includegraphics[width=\columnwidth]{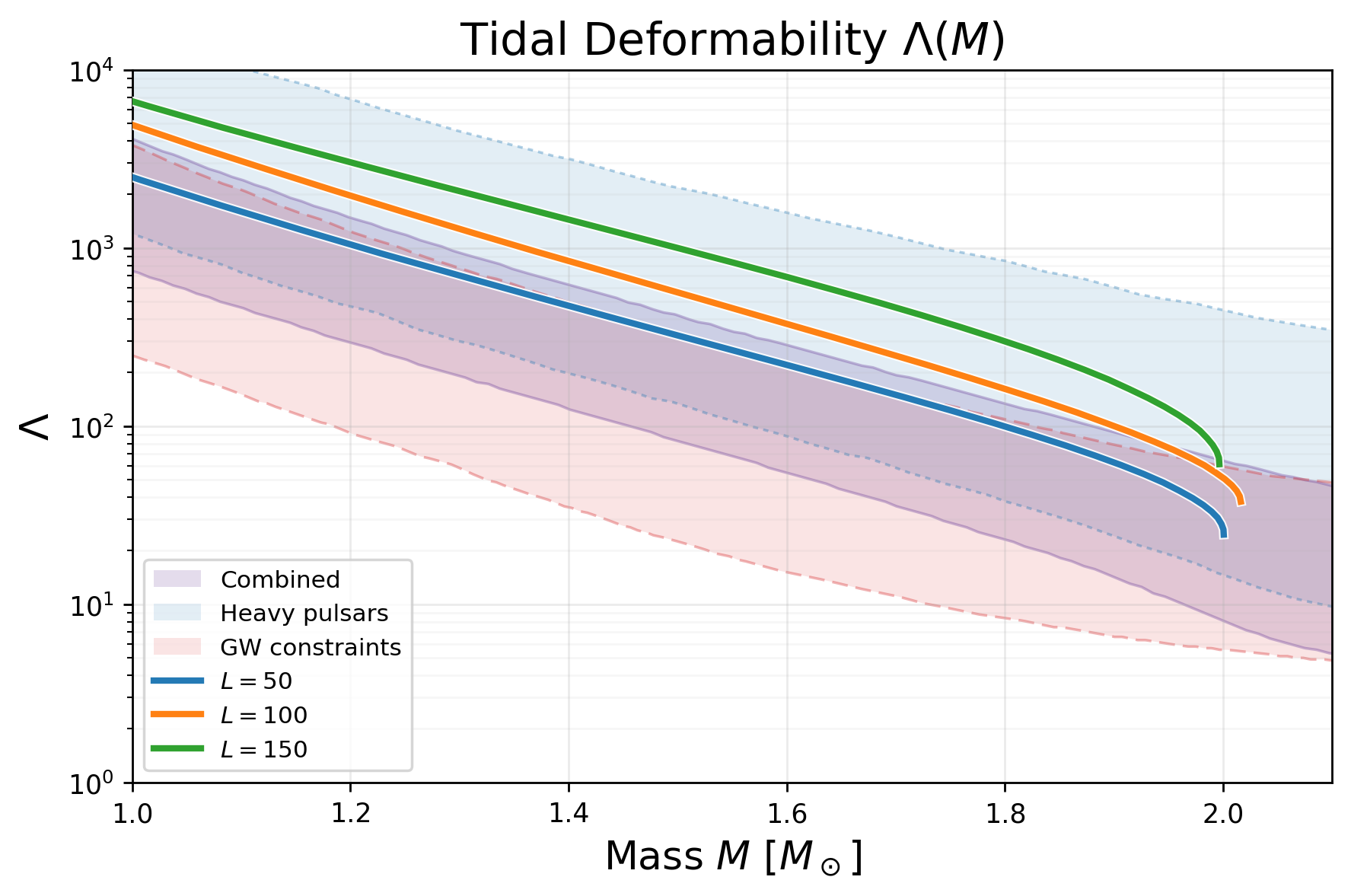}
\caption{\label{fig:tidal_deform}
Dimensionless tidal deformability $\Lambda(M)$ for cold ($s=0$) neutron-star sequences in our EOS family. The shaded regions denote the 90\% posterior-predictive EOS constraints from the nonparametric hierarchical analysis summarized in \cite{Chatziioannou_2020}. They are not direct measurements of individual $\Lambda(M)$ relations: current GW data primarily constrain binary tidal combinations at the measured chirp mass, and an EOS inference is required to map these constraints into the $\Lambda$--$M$ plane. Within our model family, all values of $L$ shown are compatible with the heavy-pulsar constraint, while the combined GW+heavy-pulsar band suggests the approximate model-dependent bound $L \lesssim 80~\mathrm{MeV}$.}

\end{figure}

We show the cold ($s=0$) mass--radius relations and tidal deformabilities predicted by our model in Figs.~\ref{fig:MR_L} and \ref{fig:tidal_deform}. These results are compared with macroscopic neutron-star constraints from massive pulsars, NICER radius measurements, and gravitational-wave observations. In the tidal-deformability case, the shaded regions should be interpreted as posterior-predictive EOS constraints obtained by mapping the GW and pulsar data through an EOS inference, rather than as direct measurements of an individual $\Lambda(M)$ relation. Current GW observations primarily constrain binary tidal combinations at the measured chirp mass, and an EOS assumption is required to translate these measurements into the $\Lambda$--$M$ plane. Within our one-parameter EOS family, increasing $L$ generally increases the radii and tidal deformabilities of canonical-mass stars. Comparison with the combined GW+heavy-pulsar posterior-predictive band therefore suggests an approximate, model-dependent consistency criterion $L \lesssim 80~\mathrm{MeV}$.

\begin{figure}[t]
\includegraphics[width=\columnwidth]{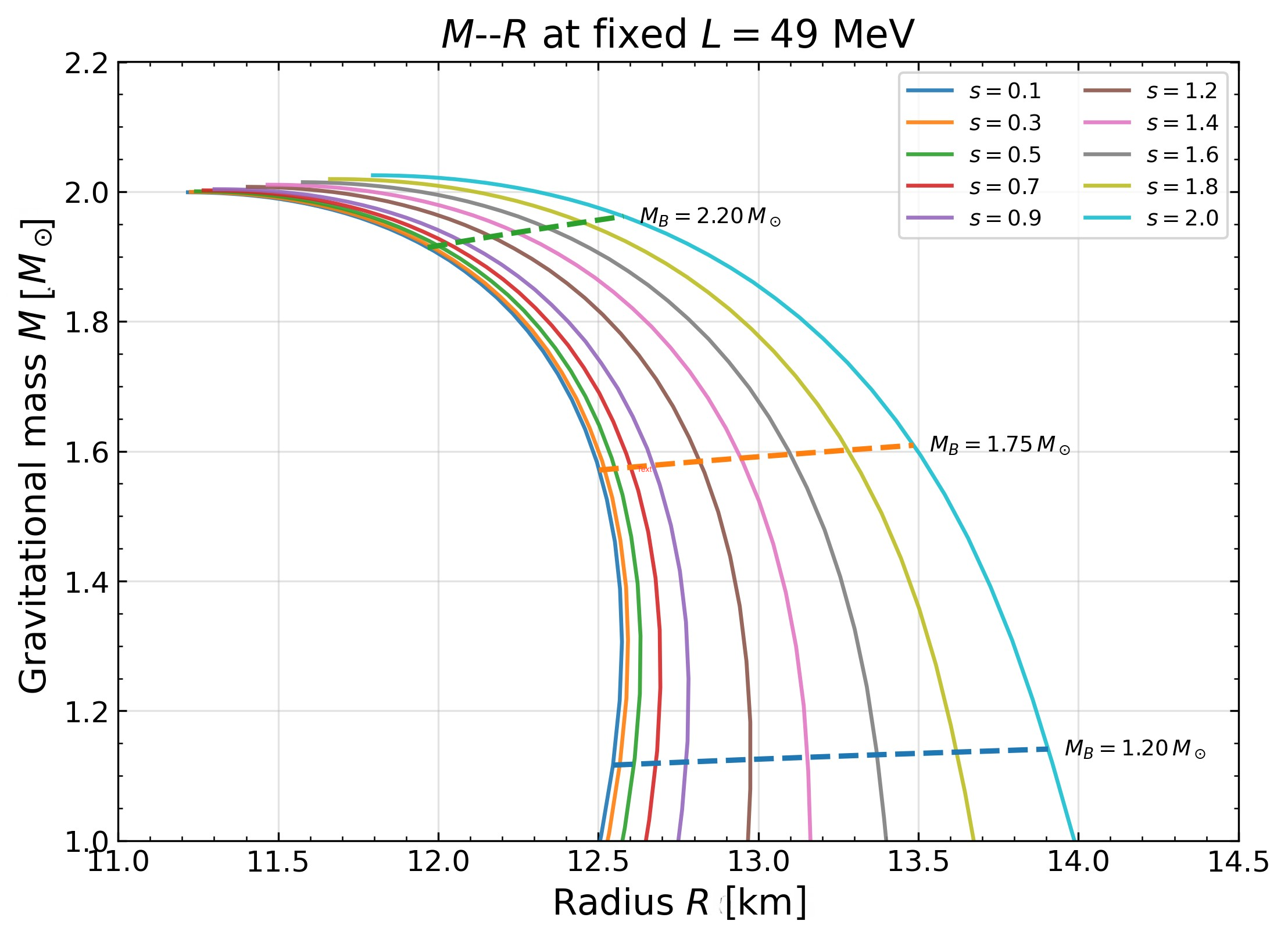}
\caption{\label{fig:MR_curve_T} Mass--radius relations for several values of the entropy per baryon. The added thermal pressure leads to higher entropy per baryon stars having larger radii at a given mass. The dashed lines track the lines of constant baryon mass $M_B = am_n$ where a is the total baryon number of the star.} 
\end{figure}

We extend this analysis to finite temperature by constructing mass--radius relations for configurations at fixed entropy per baryon, as shown in Fig \ref{fig:MR_curve_T}. We also present evolutionary tracks at constant baryon mass, which is approximately conserved during the early evolution of proto--neutron stars in the absence of significant mass loss. Increasing entropy provides additional thermal pressure support, shifting configurations toward larger radii and higher gravitational masses for a given baryon mass. Consequently, the binding energy $M_B$-$M_G$ decreases with increasing entropy. As the star cools, the loss of thermal support leads to a reduction in both radius and gravitational mass at fixed baryon mass, and the star becomes more tightly bound.

\section{$g$-MODE OSCILLATIONS}
\label{gmodesection}

In this section, we  outline the equations needed in order to solve for the g-mode frequencies. When a parcel of fluid is displaced in a stratified medium, buoyancy acts as a restoring force, leading to oscillations about its equilibrium position. This oscillation is described by the displacement vector field $\xi_{nlm}(\vec{r})e^{i\omega t}$ where $n,l,m$ are the radial, polar, and azimuthal mode indices. The spatial part of the displacement vector is expanded as: 
\begin{equation}
    \xi^i_l = \frac{1}{r} W_l Y_l^m \delta^i_r + \frac{1}{r^2}V_l \partial_\theta Y^m_l \delta^i_\theta  + \frac{1}{r^2 \sin^2{\theta}} V_l Y_l^m \delta^i_\phi
\end{equation}
Where the radial ($W_l$) and angular ($V_l$) amplitudes are functions of radius only. To simplify the analysis, we adopt the relativistic Cowling approximation, in which metric perturbations are neglected - an approximation known to reproduce full general relativistic results within $\sim5$ – $10\%$ \cite{Zhao_2022}. Under this approximation, the linearized perturbation equations for cold neutron stars take the form~\cite{Jaikumar2021, Constantinou_2021,Tran2023}

\begin{equation}
\label{oscillations}
    \begin{aligned}
        \frac{1}{r^2e^{\lambda/2}}\partial_{r}(r^2e^{\lambda/2}\xi^r) - \frac{\delta p_l e^\nu}{(p + \epsilon)r^2\omega_n^2}l(l+1) + \frac{\Delta p}{\Gamma p}\\
        e^{\lambda - \nu}( N^2-\omega_n^2 )h\xi^r  + \partial_r\delta p+  g(1+ \frac{1}{c_{\rm ad}^2} )\delta p = 0
    \end{aligned}
\end{equation}

These equations involve a set of thermodynamic variables that depend explicitly on the chosen EOS , including the specific enthalpy of the fluid, \( h = \epsilon + p \), and the local gravitational acceleration, \( g = - \nabla \phi =  -\nabla p / h \). The Eulerian $\delta p$ and Lagrangian $\Delta p$ variations of pressure are related by $\Delta p = \delta p + \xi \cdot \nabla p $. The g-mode oscillations depend on two characteristic sound speeds which are the equilibrium sound speed:
\begin{equation}
\label{c_e}
    c_e^2= \frac{dp}{d\epsilon}
\end{equation}
and the adiabatic sound speed:
\begin{equation}
\label{cs}
    c_{\rm ad}^2= \left (\frac{\partial p}{\partial \epsilon}\right)_{s,x_p}
\end{equation}
where the derivative is taken at constant entropy per baryon \(s\) and fixed composition. In the context of $npe$ matter, fixing the composition corresponds to holding the proton fraction \(x_p = n_p/n_B\) constant. This assumption is valid when the oscillation timescale is much shorter than the timescale for chemical  re-equilibration.
The difference between the two sound speeds gives rise to a buoyant restoring force, characterized by the Brunt--Väisälä frequency, \(N\), defined as  

\begin{equation}
    N^2 \equiv g^2 \left ( \frac{1}{c_e^2} - \frac{1}{c_s^2} \right )e^{\nu - \lambda}
\end{equation}

where the condition for stable oscillations requires \(N^2 > 0\), and we \(N = 0\) within the crust. For ease of integration the oscillation equations are typically recast with the variable transformations: $U = r^2e^{\lambda/2}\xi, \quad V = \delta p /(\epsilon + p )$ 
\begin{equation}
\label{eq:complex_U}
\frac{dU}{dr}
=
\frac{g}{c_{\rm ad}^2}\,U
+
e^{\lambda/2}
\left[
    \frac{\ell(\ell+1)e^\nu}{\omega^2}
    -
    \frac{r^2}{c_{\rm ad}^2}
\right]V ,
\end{equation}

\begin{equation}
\label{eq:complex_V}
\frac{dV}{dr}
=
\frac{e^{\lambda/2-\nu}}{r^2}
\left(\omega^2-N^2\right)U
+
g
\left(
    \frac{1}{c_{\rm eq}^2}
    -
    \frac{1}{c_{\rm ad}^2}
\right)V .
\end{equation}
At finite temperature, weak interaction rates become comparable to the oscillation timescale at which point the frozen composition approximation is no longer valid. Within the fast reaction regime one must then use a complex dynamical sound speed as introduced by \citet{Zhao_2025}:

\begin{equation}
    c_{\rm dy}^2 = c^2_{\rm eq} + \frac{c_{\rm ad}^2 - c^2_{\rm eq}}{1+ \frac{\gamma}{i\omega}}
\end{equation}
where $\omega = \omega_R + i\omega_I$ is the new complex frequency and $\gamma$ is the local beta equilibration rate defined by: 
(\cite{Zhao_2025}):
\begin{equation}
\gamma \equiv
\frac{\lambda_{n\leftrightarrow p}}{n_B}
\left(
\frac{\partial \tilde \mu}{\partial \delta x}
\right)_{n_B,s}.
\label{eq:relax}
\end{equation}
with 
\begin{equation}
\lambda_{n\leftrightarrow p}(T,n_B)
\equiv
\left.
\frac{\partial\left(\Gamma_{n\rightarrow p}-\Gamma_{p\rightarrow n}\right)}
{\partial \delta\mu}
\right|_{\delta\mu=0}.
\end{equation}
and $\Gamma$ the weak reaction rates. These rates are calculated from Eq. (A3) and (A5) of \cite{Most_2024}. The real and imaginary part of the sound speed difference are then:

\begin{equation}
\label{real_cdyn}
{\rm Re}\!\left[c_{\rm dy}^{2}\right]
=
c_{\rm eq}^{2}
+
\left(c_{\rm ad}^{2}-c_{\rm eq}^{2}\right)
\frac{\omega_R^{2}+\omega_I(\omega_I-\gamma)}
{\omega_R^{2}+(\omega_I-\gamma)^{2}} .
\end{equation}

\begin{equation}
\label{imag_cdyn}
{\rm Im}\!\left[c_{\rm dy}^{2}\right]
=
\left(c_{\rm ad}^{2}-c_{\rm eq}^{2}\right)
\frac{\omega_R\gamma}
{\omega_R^{2}+(\omega_I-\gamma)^{2}} .
\end{equation}

In the slow-reaction limit, $\gamma\ll|\omega|$, where
$|\omega|=(\omega_R^2+\omega_I^2)^{1/2}$, the dynamical sound speed approaches the frozen-composition sound speed:
\begin{equation}
{\rm Re}\!\left[c_{\rm dyn}^2\right]\rightarrow c_s^2,
\qquad
{\rm Im}\!\left[c_{\rm dyn}^2\right]\rightarrow 0.
\end{equation}
The composition of a displaced fluid element therefore remains effectively frozen. In the fast-reaction limit, $\gamma\gg|\omega|$, chemical re-equilibration occurs on a timescale much shorter than the characteristic mode timescale, giving
\begin{equation}
{\rm Re}\!\left[c_{\rm dyn}^2\right]\rightarrow c_{\rm eq}^2,
\qquad
{\rm Im}\!\left[c_{\rm dyn}^2\right]\rightarrow 0.
\end{equation}
The fluid consequently remains close to chemical equilibrium, suppressing the composition-driven buoyancy. The imaginary part of the dynamical sound speed is therefore appreciable only in the intermediate regime, where the chemical-equilibration and mode timescales are comparable, $\gamma\sim|\omega|$. Accordingly, in the finite-rate perturbation equations we make the replacement
\begin{equation}
\label{complex_replacements}
    c_{\rm ad}^2\longrightarrow c_{\rm dyn}^2(\omega),
    \qquad
    N_{\rm ad}^2\longrightarrow
    N_{\rm dyn}^2(\omega),
\end{equation}
where
\begin{equation}
\label{dynamic_brunt}
    N_{\rm dyn}^2(r;\omega)
    =
    g^2
    \left[
        \frac{1}{c_{\rm eq}^2(r)}
        -
        \frac{1}{c_{\rm dyn}^2(r;\omega)}
    \right]
    e^{\nu-\lambda}.
\end{equation}
Both \(c_{\rm dyn}^2\) and \(N_{\rm dyn}^2\) are generally complex.

The sensitivity of the $g$-mode spectrum to the symmetry energy enters explicitly through the density dependence of the equilibrium composition and, consequently, the buoyancy frequency. In beta equilibrium the proton fraction is determined by Eq ~\ref{charge} which, to leading order in the isospin asymmetry gives:
\begin{equation}
\mu_n - \mu_p \simeq 4 S(n_B)\left(1 - 2x_p\right),
\end{equation}
so that the equilibrium proton fraction $x_p = x_p(n_B)$ is governed by the density dependence of the symmetry energy $S(n_B)$. In terms of the the density-dependent symmetry-energy slope $L(n_B)$, and curvature
\begin{equation}
K_{\rm sym}(n_B) \equiv 9 n_B^2 \frac{d^2 S}{dn_B^2},
\end{equation}
using the parabolic approximation yields for the Brunt--Väisälä frequency ~\cite{Wei2020Gmodes}
\begin{equation}
\begin{split}
& N \approx 
2\left(\frac{g}{c_e}\right)
\left(\frac{x_p}{3}\right)^{1/2} \\
& \times \frac{L(n_B) - S(n_B)}
{\sqrt{
S(n_B)
\left(
\frac{10}{9}E_0(n_B)
+ \frac{2}{3}L(n_B)
+ \frac{1}{9}K_{\rm sym}(n_B)
\right)
}} .
\end{split}
\end{equation}

The numerator reflects the role of the symmetry energy and its derivatives, $L$ and $K_{\rm sym }$, in setting the Brunt--Väisälä frequency and thus g-mode spectrum.

In this work, we fix the $g$-mode's angular degree to \(l = 2\), since this mode couples efficiently to gravitational waves. We also set the radial order to \(n = 1\), corresponding to the fundamental mode, which exhibits the largest frequency and amplitude of excitation. Finally, for non-rotating stars, the oscillation modes are degenerate in the azimuthal index \(m\).


\section{Sound Speeds}
\label{soundspeed}
Once the equation of state (EOS) and the equilibrium structure equations have been solved, the only other inputs required for the perturbation equations \ref{oscillations} are the two sound speeds, given in Eqs.~(\ref{c_e}) and (\ref{cs}). The equilibrium sound speed can be obtained straightforwardly, as it corresponds to the total derivative of pressure with respect to energy density. In contrast, the calculation of the adiabatic sound speed is more involved, as detailed in this section. The adiabatic sound speed, defined in Eq.~(\ref{cs}), can be expressed as

\begin{equation}
\label{cs2}
    c_{\rm ad}^2= \left ( \frac{\partial p}{\partial n_B}\right )_{s,x_p} \bigg /\left ( \frac{\partial \epsilon}{\partial n_B}\right )_{s,x_p} 
\end{equation}
In terms of the entropy per baryon, s, and the particle fractions, $x_i$ the first law of thermodynamics reads:
\begin{equation}
    d\epsilon = Tn_Bds  + sTdn_B + \sum_{i}\mu_{i}n_Bdx_i + \sum_{i}\mu_{i}x_idn_B
\end{equation}

Since we are dealing with neutron star matter, the conditons $n_B = n_n + n_p$ and $n_e = n_p$ allow us to re-express the 1st law as:
\begin{equation}
    d\epsilon =  Tn_Bds + \frac{\epsilon + P}{n_B}dn_B  + n_B\tilde\mu dx_P 
\end{equation}
\\
where $\tilde\mu = \mu_p+\mu_e -\mu_n$. Using this to take the partial derivative of $\epsilon$ at fixed $s$ and $x_p$ leads to:
\begin{equation}
\label{dedn}
    \left ( \frac{\partial \epsilon}{\partial n_B}\right )_{s,x_p} =  \frac{\epsilon + P}{n_B}
\end{equation}
Using the Gibbs-Duhem relation to calculate the partial derrivative of the pressure we have:
\begin{equation}
\label{dpdn}
    \left ( \frac{\partial P}{\partial  n_B} \right )_{s,x_p}= sn_B\left( \frac{\partial T}{\partial n_B} \right )_{s,x_p}  + \sum_{i}n_Bx_i\left( \frac{\partial \mu_i}{\partial n_B} \right )_{s,x_p}
\end{equation}

Plugging in Eq (\ref{dedn}), (\ref{dpdn}) into [\ref{cs2}] we get the final expression for the adiabatic sound speed:

\begin{equation}
\label{cs_squared}
    c_{\rm ad}^2 = \frac{n_B}{\epsilon + P } \left [sn_B\left( \frac{\partial T}{\partial n_B} \right )_{s,x_p}  + \sum_{i}n_Bx_i\left( \frac{\partial \mu_i}{\partial n_B} \right )_{s,x_p} \right ] 
\end{equation}

We now have 4 unknowns: the derivatives of the chemical potentials and the derivatives of the temperature. In what proceeds we will denote partial derivatives with $x_p$ and $s$ held constant by a prime.  Motivated by \cite{Tran2023, Shirke2025} we solve for these unknowns by taking the derivative of the scalar field equation(\ref{sigma_equation}) :
\begin{widetext}
\begin{equation}
\label{y_deriv}
    \begin{split}
         &y'\left [ -2y + \frac{4by}{m^2c_{\omega}}(1-y^2)  - \frac{6yc}{m^4c^2_{\omega}} (1-y^2)^2 -\frac{8c_{\sigma}c_{\omega}n_{B}^2}{m^2y^5}  \right]  +\frac{4c_{\sigma}c_{\omega}n_{B}}{m^2y^4}
         +\frac{4 \eta c_{\sigma}c_{\rho}m_\rho^2 \rho \dot{\rho}}{c_\omega m^2}\\
         & - \frac{2c_{\sigma}}{\pi^2}\int_{0}^{\infty} dk \left [ \frac{-y'mm^*k^2}{(k^2 + m^{*2})^{3/2}}(f + \bar{f}) + \frac{k^2}{\sqrt{k^2+m^{*2}}}\left ( f' + \bar{f}' \right ) \right]
    \end{split}
\end{equation}
\end{widetext}

In terms of $E_{k}^* = \sqrt{k^2+{m^*}^2}$ the derivatives of the Fermi-Dirac distributions, $f'$ are given by:
\begin{equation}
   f'_i = -f_i(1-f_i)\left [ \left ( \frac{mm^*y'}{E^*_{k_i}} -\nu'_i \right )\frac{1}{T} \right . - \left .\left ( \frac{E^*_{k_i}- \nu_i}{T^2} \right )T' \right ]
\end{equation}

The rho-field equation becomes: 
\begin{equation}
\label{rho_deriv}
     \left [ 1- \eta(1-y^2)\frac{c_\rho}{c_\omega}\right ] \rho ' + \frac{2c_\rho\eta yy' \rho}{c_\omega}  = \frac{g_{\rho}}{2m^2_{\rho}}(2x_p-1)
\end{equation}

We now have 6 unknown derivatives: $y',\,  \rho'\, , \mu_n' \, , \mu_p' \, , \mu_e'\, ,T' $. To solve for these, we take derivatives of the constraint equations of proton fraction:
\begin{equation}
\label{xp_deriv}
        x_pn_B = n_p \rightarrow x_p = \frac{1}{\pi^2}\int_0^\infty k^2dk\left ( f_p' - \bar{f_p'}\right )
\end{equation}

\noindent charge neutrality
\begin{equation}
\label{charge_deriv}
    \begin{split}
        &n_e = n_p\\
        &\rightarrow \frac{1}{\pi^2}\int_0^\infty k^2dk\left ( f_e' - \bar{f_e'}\right ) = \frac{1}{\pi^2}\int_0^\infty k^2dk\left ( f_p' - \bar{f_p'}\right )
    \end{split}
\end{equation}
baryon density 
\begin{equation}
\label{baryon_deriv}
    \begin{split}
        &n_B = n_p + n_n \\ &\rightarrow 1 = \frac{1}{\pi^2}\int_0^\infty k^2dk\left ( f_p' - \bar{f_p'}\right ) + \frac{1}{\pi^2}\int_0^\infty k^2dk\left ( f_n' - \bar{f_n'}\right )
    \end{split}
\end{equation}
For our final equation, we take a derivative of the entropy per baryon leading to :
\begin{equation}
\label{entropy_deriv}
\begin{split}  
s
=
\sum_i \frac{1}{\pi^2}
\int_0^\infty dk\, k^2
\left[
f_i'
\left(\frac{E^*_{i,k}-\nu_i}{T}\right)
+
\bar f_i'
\left(\frac{E^*_{i,k}+\nu_i}{T}\right)
\right].
\end{split}
\end{equation}
We know have our 6 unknowns $y', \rho', \mu_n', \mu_p', \mu_e',T'$ and the 6 equations (\ref{y_deriv})(\ref{rho_deriv})-(\ref{entropy_deriv}) which can be solved simultaneously for a given $n_B$.The derivative of $\tilde \mu$ in Eq.(\ref{eq:relax}) is found in a similar way to the adiabatic sound speed by taking derivatives of the field equations except with the baryon density and entropy per baryon being held fixed.\\

\section{Results}
\label{symmetry}

\begin{figure}[t]
\includegraphics[width=\columnwidth]{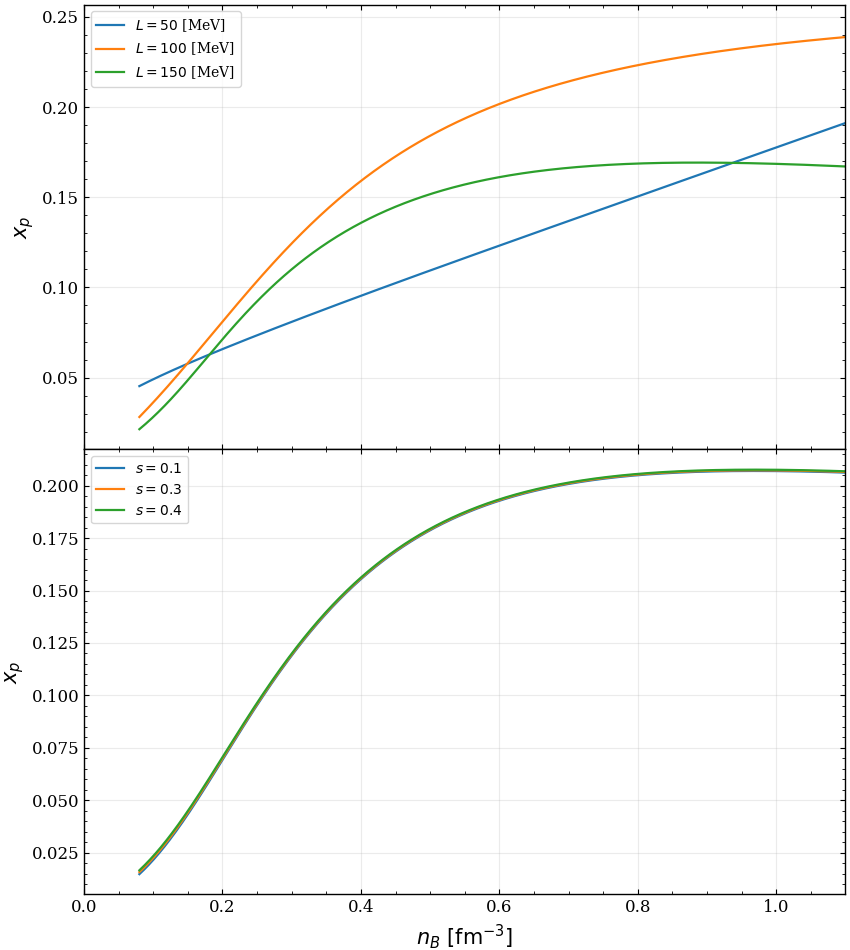}
\caption{\label{fig:comp_profiles} Top: beta-equilibrium proton fraction $x_p$ as a function of baryon density for several values of the symmetry energy slope $L$ at $T = 0$. Larger $L$ steepens the rise of $x_p$ at low densities, while at higher densities the trend reverses: $x_p$ flattens, whereas the $L = 50$ MeV EOS continues to grow nearly linearly. Bottom: proton fraction for several values of the entropy per baryon $s$ at fixed $L = 105$ MeV. Finite entropy leaves $x_p$ essentially unchanged in the regime of interest. These composition gradients underlie the behavior of the sound-speed difference $c_{\rm s}^2 - c_{\rm eq}^2$ and the g-mode frequencies. }
\end{figure}

\begin{figure}[h]
\includegraphics[width=\columnwidth]{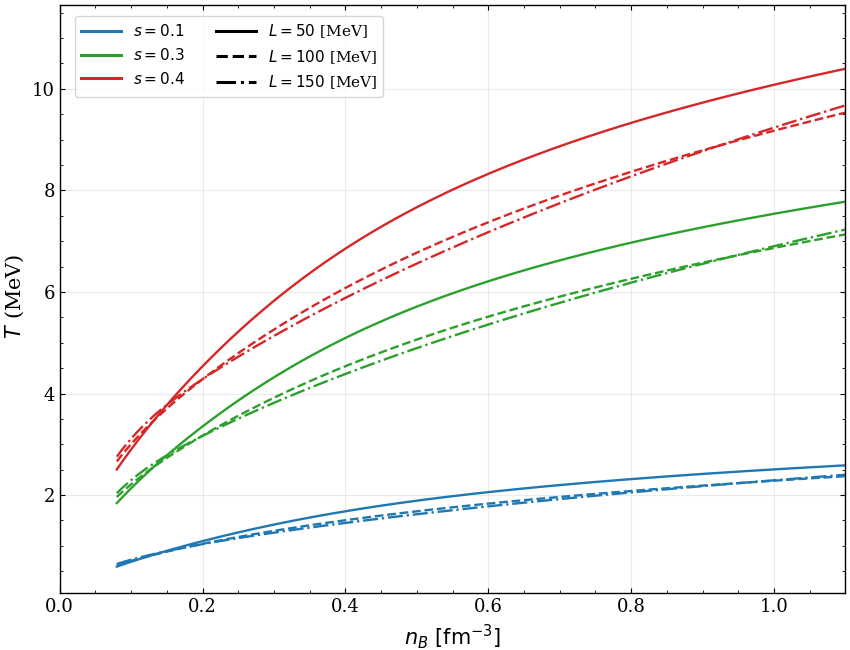}
\caption{\label{fig:temp_profiles} Temperature profiles $T(n_{\rm B})$ at fixed entropy per baryon $s = 0.1$, $0.3$, and $0.4$ for EOSs with symmetry energy slope $L = 50$, $100$, and $150$ MeV. The temperature increases monotonically with density and scales strongly with $s$, reaching $T \approx 10$ MeV in the inner core for $s = 0.4$. The dependence on $L$ is comparatively mild: the profiles are nearly $L$-independent at low densities, while at higher densities the $L = 50$ EOS yields temperatures a few MeV above the $L = 100$ and $150$ cases. }

\end{figure}

\begin{figure}[t]
\centering
\includegraphics[width=\columnwidth]{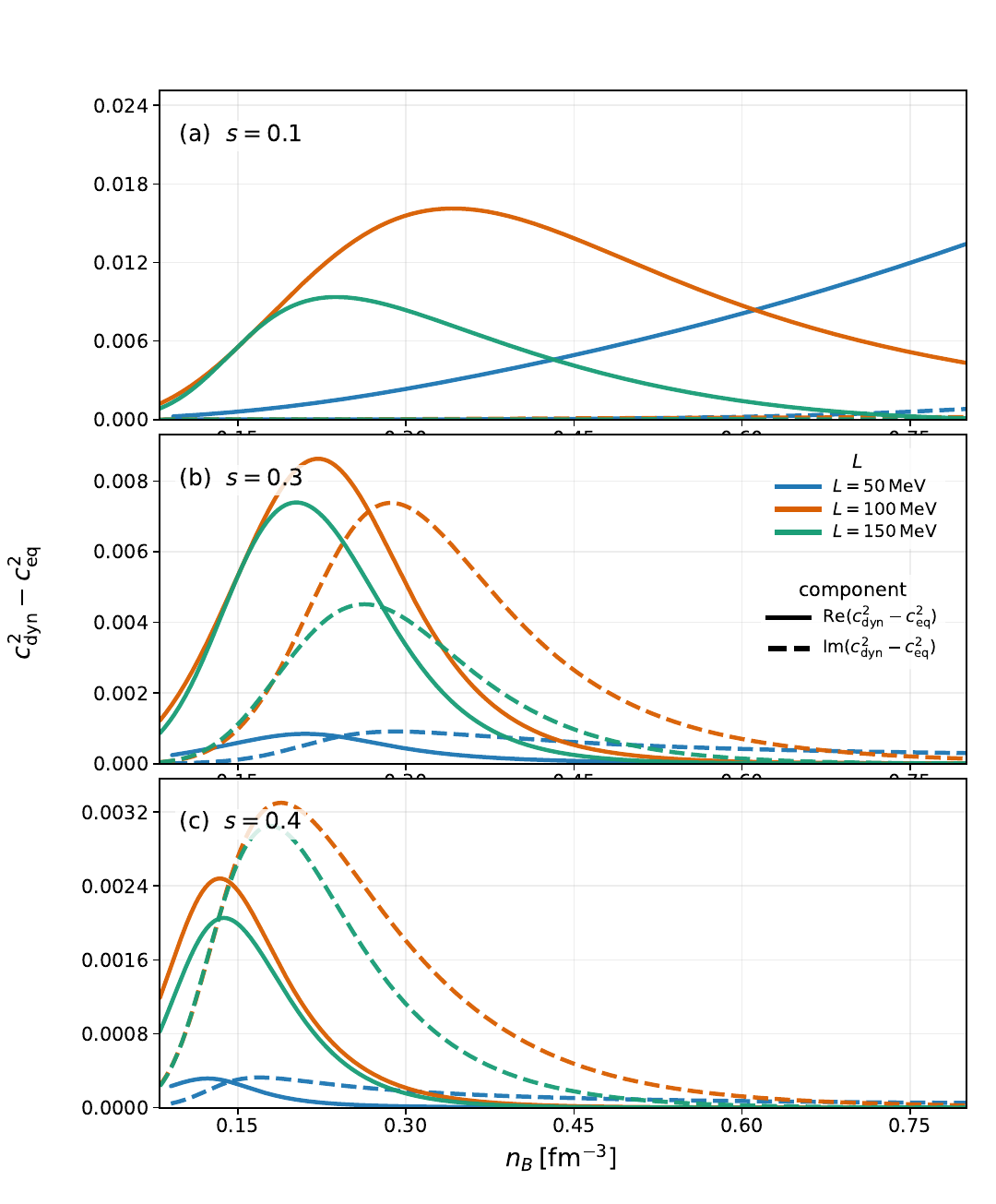}
\caption{\label{fig:cdyn_diff}
Complex dynamical sound-speed difference, $c_{\rm dyn}^2-c_{\rm eq}^2$, as a function of baryon density $n_B$ at fixed angular frequency $\omega=1000\,{\rm s}^{-1}$. The three panels correspond to entropy per baryon $s=0.1$, $0.3$, and $0.4$, while colors denote different values of the symmetry-energy slope parameter $L$. Solid and dashed curves show, respectively, the real and imaginary parts of $c_{\rm dyn}^2-c_{\rm eq}^2$.  Increasing entropy increases the weak-interaction relaxation rate and therefore strengthens the imaginary part of the dynamical sound-speed response, which governs bulk-viscous damping. As the damping becomes stronger, the real part of $c_{\rm dyn}^2-c_{\rm eq}^2$ approaches zero because $c_{\rm dyn}^2$ is driven toward the equilibrium limit $c_{\rm eq}^2$, suppressing the composition-driven buoyancy. The dependence on $L$ reflects its influence on the equilibrium composition gradient and on the underlying frozen-composition sound-speed difference.
}
\end{figure}

The density dependence of the proton fraction provides the equilibrium composition stratification that sources the $g$-mode buoyancy. This can be best shown in the low temperature, frozen-composition limit where the sound speed can be written as:
\begin{equation}
\label{eq:csce_main}
    c_{\rm ad}^2-c_{\rm eq}^2
    =
    -\frac{n_B}{\epsilon+P}
    \left(\frac{\partial P}{\partial x_p}\right)_{n_B,s}
    \frac{dx_p^\beta}{dn_B}.
\end{equation}
Whose derivation is shown in its entirety in Appendix ~\ref{app:sound_speed_difference}. Thus the buoyant restoring force is controlled directly by the equilibrium composition gradient $dx_p^\beta/dn_B$, together with the thermodynamic susceptibility $(\partial P/\partial x_p)_{n_B,s}$.

As shown in Fig.~\ref{fig:comp_profiles}, varying the symmetry-energy slope $L$ changes both the magnitude and shape of $x_p(n_B)$. At low densities, larger $L$ produces a more rapid increase of the proton fraction, while at higher densities the large-$L$ curves flatten while the lower-$L$ models continue
to vary more strongly with density. Since buoyancy is generated by spatial variations in the equilibrium composition, regions where $x_p(n_B)$ becomes nearly flat have little composition contrast and therefore provide a much weaker source for the $g$-mode's restoring force. Consequently, the sound-speed
difference tends toward zero at high density where the $x_p$ profile flattens, as the displaced fluid element is no longer compositionally distinct from its surroundings. Overall, increasing $L$ enhances the amplitude of the sound-speed difference over the low- and intermediate-density regions where the proton-fraction profile varies most strongly, with the largest values of $c_{\rm ad}^2-c_{\rm eq}^2$ occurring for $L\simeq 100\,{\rm MeV}$.

Thermal effects enter through the background composition, thermal
stratification, and finite-rate response of the fluid. The proton-fraction profiles in Fig.~\ref{fig:comp_profiles} show that finite entropy has no significant effect on $x_p$, while the temperature profiles in Fig.~\ref{fig:temp_profiles} show a much stronger entropy dependence throughout the core. The direct thermal contribution to the frozen-composition buoyancy enters through the first term in Eq ~\ref{cs_squared}. 
For the isentropic profiles considered here, the composition-gradient term $dx_p^\beta/dn_B$ provides the dominant source of the frozen-composition sound-speed difference. On the other hand thermal effects are primarily driven by the $T^6$ ~\cite{Most_2024} temperature dependence of the week interaction rates  which drives $c_{\rm dyn}^2$ toward $c_{\rm eq}^2$ and suppresses the effective buoyant restoring force. The temperature profile itself has only a mild effect on the g-mode through its modification of the thermodynamic susceptibility.
The strength of this thermal suppression of the difference is quantified the ratio $\gamma/\omega$ as shown in Eqs ~\ref{real_cdyn} $\&$~\ref{imag_cdyn}. As $s$ increases, the corresponding increase in $\gamma$ allows the composition of a displaced fluid element to adjust more rapidly toward chemical equilibrium. The imaginary part of $c_{\rm dyn}^2-c_{\rm eq}^2$ consequently becomes more important over the range considered. At the same time, the real part approaches zero as $c_{\rm dyn}^2$ is driven toward $c_{\rm eq}^2$, reducing the effective composition-driven buoyant restoring force. 

\begin{figure}[t]
\includegraphics[width=\columnwidth]{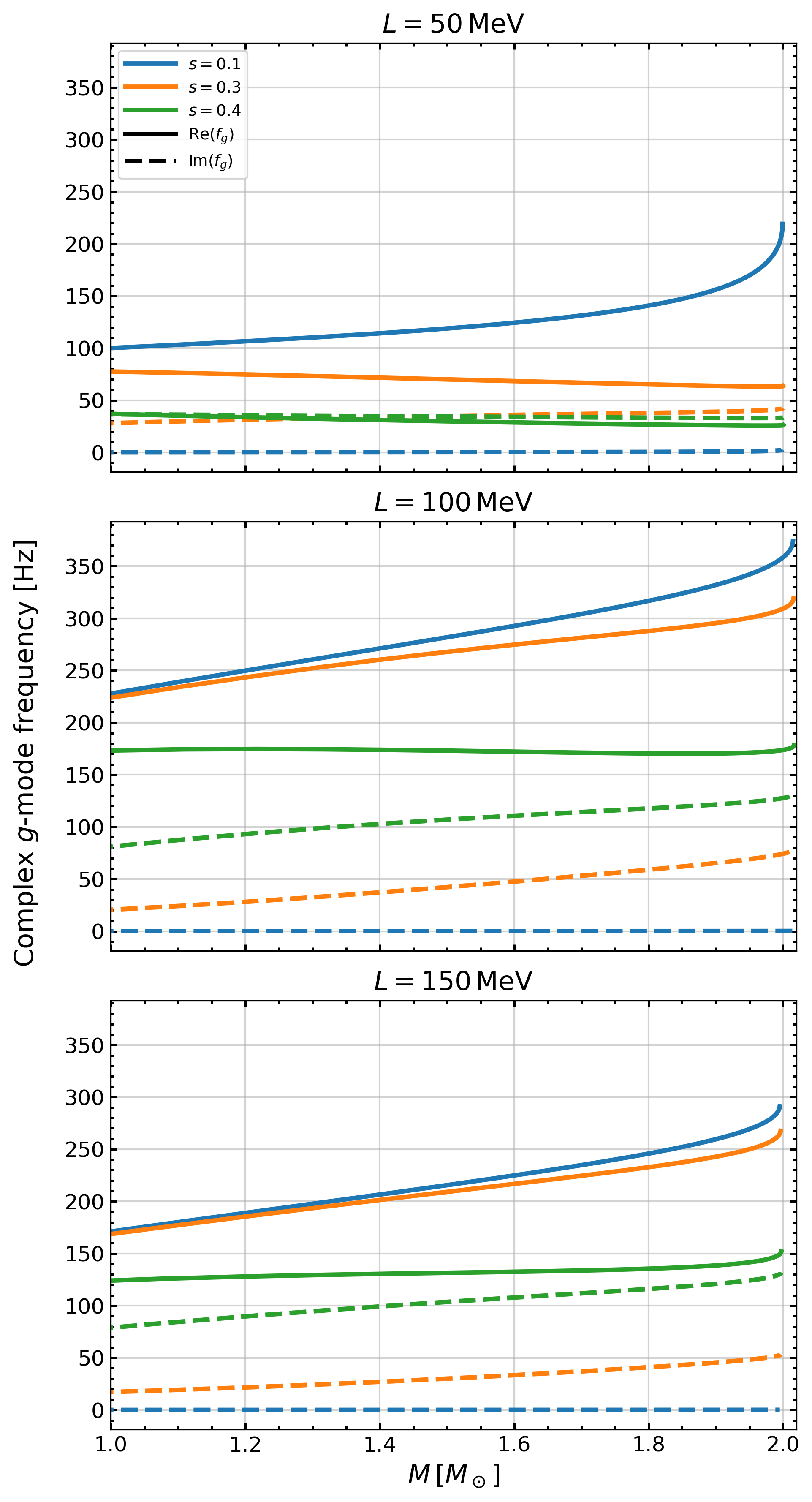}
\caption{\label{fig:gmode_freq_m}
Real and imaginary parts of the complex $g$-mode frequency as functions of gravitational mass. The three panels correspond to symmetry-energy slope parameters $L=50$, $100$, and $150\,{\rm MeV}$. Colors denote entropy per baryon $s=0.1$, $0.2$, and $0.4$. Solid curves show the oscillation frequency $\operatorname{Re}(f_g)$, while dashed curves show the damping rate $\operatorname{Im}(f_g)$. The real frequency generally increases with stellar mass, while increasing entropy weakens the buoyant restoring force and enhances bulk-viscous damping.
}
\end{figure}

\indent
We calculate the complex $g$-mode frequency,
$f_g=f_{g,r}+if_{g,i}$, as a function of gravitational mass for several values of the symmetry-energy slope $L$ and entropy per baryon $s$. The results are shown in Fig.~\ref{fig:gmode_freq_m}. The real part $f_{g,r}$ determines the oscillation frequency, while the imaginary part $f_{g,i}$ determines the damping rate. For a fixed $L$, the real part generally increases with stellar mass. Increasing the entropy reduces the real frequency by weakening the effective buoyant restoring force and simultaneously increases the importance of the imaginary component associated with bulk-viscous damping.

\begin{figure*}[t]
\includegraphics[width=\textwidth]{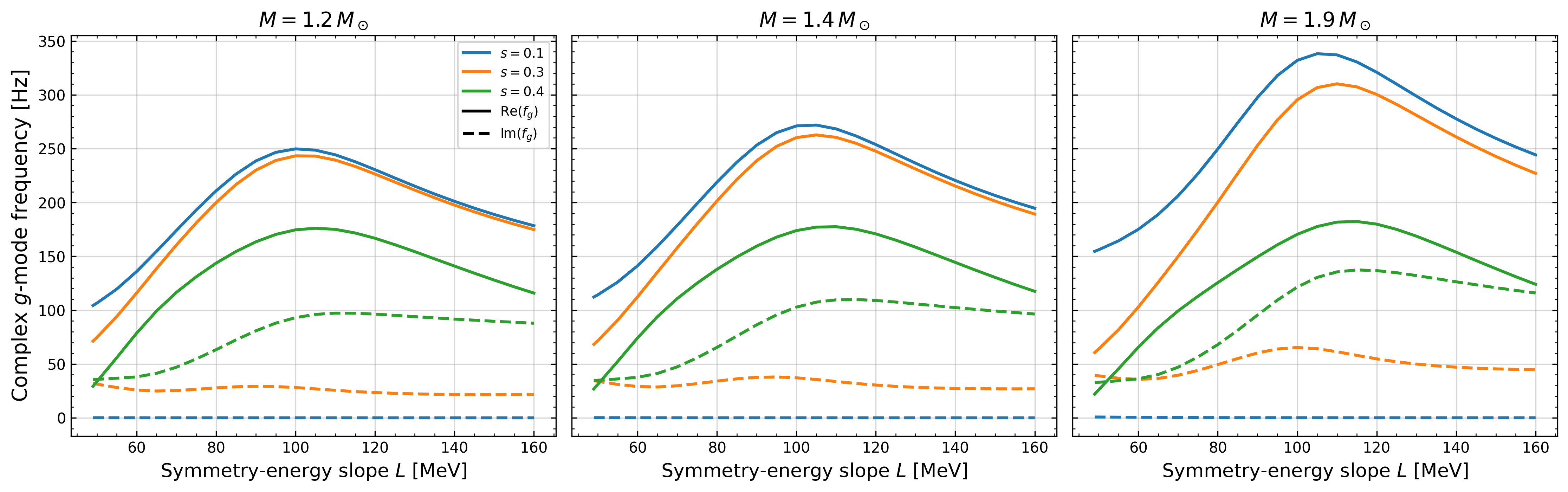}
\caption{\label{fig:gmodes_L}
Real and imaginary parts of the complex $g$-mode frequency as functions of the symmetry-energy slope parameter $L$. The three panels correspond to stellar masses $M=1.2$, $1.5$, and $1.9\,M_\odot$, while colors denote different values of the entropy per baryon. Solid and dashed curves show, respectively, $\operatorname{Re}(f_g)$ and $\operatorname{Im}(f_g)$. At fixed entropy, the real frequency is largest at intermediate $L$ and lower at both small and large $L$, reflecting the non-monotonic dependence of the equilibrium composition gradient and buoyant restoring force on the symmetry energy. Increasing entropy reduces the real oscillation frequency and increases the relative importance of bulk-viscous damping.
}
\end{figure*}

The real part of the $g$-mode frequency is governed primarily by the surviving real component of
$c_{\rm dyn}^2-c_{\rm eq}^2$, which sets the buoyant restoring force through the underlying composition gradients. As shown in Fig.~\ref{fig:gmodes_L}, this connection manifests in a non-monotonic dependence of $\operatorname{Re}(f_g)$ on $L$ for all stellar masses considered: at low $L$, the frequency increases with $L$, reaches a maximum around $L \approx 100$--$110~\mathrm{MeV}$ depending on the mass, and decreases thereafter. This mirrors the behavior of the composition gradients, which attain their largest amplitudes at similar values of $L$ which maximizes the buoyancy that supports the mode. These results are consistent with Ref.~\cite{Sun_2025}, which finds that the real g-mode frequency increases with $L$ over the more limited range considered there; our analysis extends to larger $L$ and reveals the subsequent turnover.

\begin{figure*}[t]
\includegraphics[width=\textwidth]{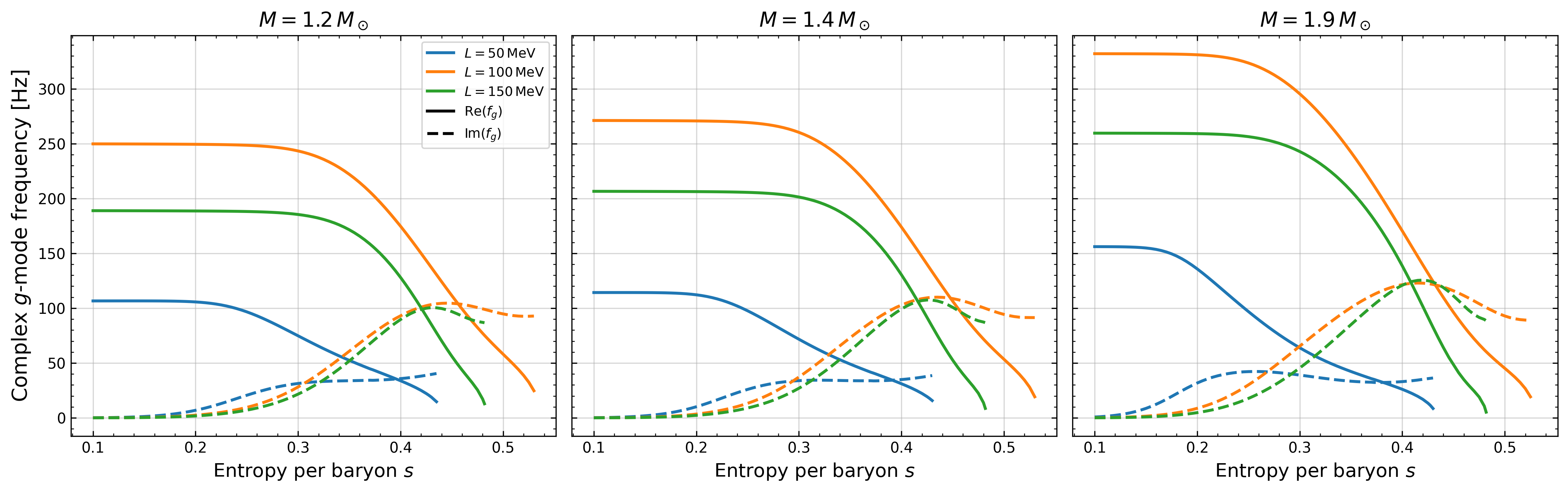}
\caption{\label{fig:freq_v_s}
Real and imaginary parts of the complex $g$-mode frequency as functions of entropy per baryon $s$. The three panels correspond to stellar masses $M=1.2$, $1.5$, and $1.9\,M_\odot$, while colors denote different values of the symmetry-energy slope parameter $L$. Solid curves show $\operatorname{Re}(f_g)$ and dashed curves show $\operatorname{Im}(f_g)$. The imaginary component reaches a maximum near, \(\omega\sim\gamma\), and decreases in the fast-reaction regime, \(\omega\ll\gamma\).
}

\end{figure*}

At finite temperature, this picture is modified by bulk-viscous damping, as shown in Figs.~\ref{fig:gmode_freq_m}--\ref{fig:freq_v_s}. Increasing the entropy per baryon raises the beta-equilibration rate $\gamma$ and drives the dynamical response toward the equilibrium sound-speed limit. Consequently, the effective buoyant restoring force and $\operatorname{Re}(f_g)$ decrease. Thermal re-equilibration also flattens the dependence of $\operatorname{Re}(f_g)$ on $L$, thereby reducing the sensitivity of the observable oscillation frequency to the symmetry energy.

As shown in Fig.~\ref{fig:freq_v_s}, the damping
rate $\operatorname{Im}(f_g)$ exhibits a non-monotonic dependence on entropy for the modes that pass through the relaxation crossover. It initially increases as the weak-interaction rate approaches the characteristic mode frequency, reaches a maximum when the chemical-equilibration and mode timescales become comparable, and subsequently decreases in the fast-reaction regime. This behavior reflects the corresponding dissipative part of the dynamical sound-speed response, which vanishes in both the slow and fast-reaction limits and is largest in the intermediate regime, $\gamma\sim|\omega|$ as shown by Eq ~\ref{imag_cdyn}. Because $\gamma$ varies throughout the stellar core and the complex eigenfrequency changes with entropy, the maximum of $\operatorname{Im}(f_g)$ represents a global relaxation crossover rather than the exact condition $\gamma=|\omega|$ at every radius.

\begin{figure*}[t]
\includegraphics[width=\textwidth]{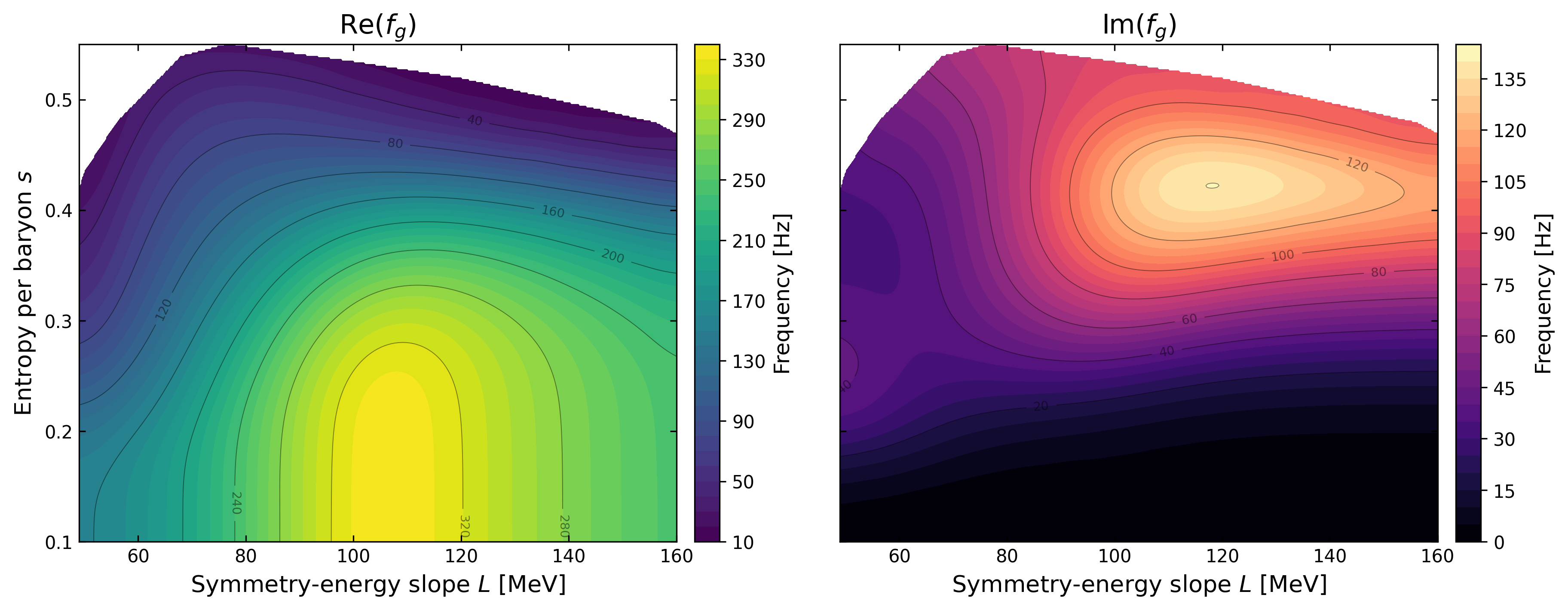}
\caption{\label{fig:heat_map}
Real (left) and imaginary (right) components of the complex (g)-mode frequency at a fixed gravitational mass of $M=1.9\,M_\odot$, shown as functions of the symmetry-energy slope (L) and entropy per baryon (s).Colors denote the frequency in hertz, and the black curves show labeled isofrequency contours. Uncolored regions indicate parameter combinations for which sufficient valid (g)-mode solutions were unavailable.
}

\end{figure*}

The complex $g$-mode frequency therefore reflects the combined influence of the symmetry-energy slope and the thermal state of the star. Neither can be inferred independently without information about the other. This is shown explicitly Fig ~\ref{fig:heat_map} in which we plot the g-mode frequency in the $L-s$ plane with curves demonstrating isofrequency contours. These contours take the form of arches for $\operatorname{Re}(f_g)$ and ellipse like curves for  $\operatorname{Im}(f_g)$. Observations of oscillations in protoneutron stars or post-merger remnants would consequently constrain $L$ only up to an uncertainty associated with the stellar thermal profile and weak-interaction rate.

\section{Conclusion}
\label{conclusion}
In this work, we have investigated the impact of finite temperature on composition-driven g-mode oscillations in warm neutron stars. Employing a chiral SU(2) sigma model to describe hadronic matter and incorporating thermal effects through isentropic profiles, we systematically varied the symmetry energy slope parameter $L$, to show that temperature and nuclear microphysics play a critical and coupled role in determining the g-mode spectrum of massive neutron stars.

We find that the real part of the $g$-mode frequency $\omega_g$ exhibits a non-monotonic dependence on the symmetry energy slope $L$ across all stellar masses considered, with a peak occurring near $L \sim 100-110$ MeV. This behavior is directly linked to the underlying compositional gradients, which are governed by the density dependence of the symmetry energy. To our knowledge, the dependence of $g$-mode frequencies on a continuous and extended range of $L$ has not been studied previously. Compared to existing calculations of nuclear and neutron star properties \cite{Zhao_2025,Lopes2024,Dutra2014} that considered a few discrete values over a limited interval (47 $\lesssim L\lesssim$ 92) MeV, extending the analysis beyond ($L\sim 100$) MeV reveals a non-monotonic trend, which appears to be driven by the corresponding rise and subsequent decline of the proton fraction gradient. Given our choice of a specific EOS, the chiral $SU(2)$ model, whether this behavior is generic across RMF frameworks remains an open question and motivates systematic studies of composition gradients and $g$-mode frequencies for $L\gtrsim 100$ MeV within a broader set of models.

At finite temperature, the principal effect is the finite-rate relaxation of the composition during the oscillation. As entropy increases, the higher temperature raises the weak-interaction re-equilibration rate, driving the fluid response toward the equilibrium sound-speed limit and suppressing the effective buoyant restoring force. The complex eigenfrequency is determined by the competition between the weak-interaction relaxation rate and the mode timescale. As entropy increases, the real oscillation frequency decreases, while the imaginary component associated with bulk-viscous damping becomes increasingly important.

The symmetry-energy slope $L$ remains important because it sets the underlying composition stratification and hence the available frozen-composition buoyancy. However, thermal re-equilibration reduces this buoyancy and weakens the sensitivity of the $g$-mode spectrum to $L$ at larger entropy.

 We have only considered $n,p,e$ matter in this work. If additional baryonic degrees of freedom (eg., hyperons) or a quark-hadron phase transition occurs in the stellar core, it can introduce additional composition gradients that modify the core $g$-mode spectrum \cite{Jaikumar2021, Constantinou_2021}. In such cases, the $g$-mode frequencies are expected to depend not only on the hadronic or quark symmetry energy \cite{Chu2014QuarkSymmetryEnergy} but also on the properties of the phase transition (e.g., first order, crossover or mixed-phase construction), thereby partially shifting the sensitivity away from purely nuclear parameters.

Overall, this study highlights that g-modes provide a unique and complementary probe of dense matter physics, particularly the poorly constrained density dependence of the symmetry energy. The inclusion of finite temperature effects is essential for realistic modeling of proto-neutron stars and post-merger remnants, and may play a decisive role in interpreting future gravitational-wave observations.

We note several caveats to the present analysis. First, our modeling assumes a neutron star composed only of npe matter and one specific model for the equation of state; whereas it is probable that additional degrees of freedom such as hyperons ~\cite{Tran2023}, meson condensates ~\cite{Ma_2022}, or deconfined quarks ~\cite{Annala_2023} may appear and alter both the equation of state and the resulting oscillation spectrum. Likewise, at sufficiently high temperatures, neutrino trapping becomes important due to its reduced mean free path ~\cite{Alford_2018} and can modify the thermodynamic gradients that give rise to buoyancy, potentially altering both the behavior and existence of g-modes. 

An important direction for future work is to investigate the observability of these modes during binary neutron-star inspiral. In particular, the gravitational-wave phase shift produced by the resonant tidal excitation of a $g$-mode, following approaches such as that of Ref.~\cite{Counsell_2024, Counsell_2025}, could provide a direct connection between the mode spectrum and an observable inspiral signature. Such a study would allow the influence of the symmetry energy on the resonance frequency, tidal coupling, and  phase shift to be quantified. Extending the present analysis to include resonant tidal excitation, gravitational-wave phase evolution, neutrino trapping, and a broader range of microscopic equations of state is therefore left for future work.

\section{Acknowledgement}
 P.J. is supported by the U.S. National Science Foundation Grant PHY-2310003. D.M.Z. acknowledges support from the Google Summer Research Assistantship.

\appendix

\section{Real and imaginary parts of the complex mode equations}
\label{app:complex_modes}

Finite-rate chemical equilibration makes the dynamical adiabatic
index, the buoyancy frequency, the perturbation variables, and the
eigenfrequency complex. We therefore write
\begin{equation}
    U=U_r+iU_i,\qquad
    V=V_r+iV_i,\qquad
    \omega=\omega_r+i\omega_i .
\end{equation}
We adopt perturbations proportional to \(e^{i\omega t}\), so that
\(\omega_i>0\) corresponds to damping.

The complex dynamical adiabatic index is
\begin{equation}
\label{eq:Gamma_complex}
\begin{split}
    \Gamma
    &=
    \Gamma_{\rm eq}
    +
    \frac{\Gamma_{\rm ad}-\Gamma_{\rm eq}}
         {1-i\gamma/\omega}
=
    \Gamma_{\rm eq}
    +
    (\Gamma_{\rm ad}-\Gamma_{\rm eq})
    \frac{
        \omega_r^2+\omega_i(\omega_i-\gamma)
    }{
        \omega_r^2+(\omega_i-\gamma)^2
    }
    +i(\Gamma_{\rm ad}-\Gamma_{\rm eq})
    \frac{
        \omega_r\gamma
    }{
        \omega_r^2+(\omega_i-\gamma)^2
    }
    \equiv \Gamma_r+i\Gamma_i .
\end{split}
\tag{B1}
\end{equation}

The corresponding complex Brunt--V\"ais\"al\"a frequency is written as
\begin{equation}
\label{eq:N2_complex}
\begin{split}
    N^2
    &=
    g^2\frac{\epsilon+p}{p}
    \left(
        \frac{1}{\Gamma_{\rm eq}}
        -
        \frac{\Gamma_r}{\Gamma_r^2+\Gamma_i^2}
    \right)e^{\nu-\lambda}
    +i\,g^2\frac{\epsilon+p}{p}
    \left(
        \frac{\Gamma_i}{\Gamma_r^2+\Gamma_i^2}
    \right)e^{\nu-\lambda}
    \equiv N_r+iN_i .
\end{split}
\tag{B2}
\end{equation}

Splitting the two complex perturbation equations into their real and
imaginary parts gives four coupled real first-order equations:

\begin{equation}
\label{eq:dUr}
\begin{split}
\frac{dU_r}{dr}
={}&
\frac{(\epsilon+p)g}
     {p(\Gamma_r^2+\Gamma_i^2)}
     \left(\Gamma_rU_r+\Gamma_iU_i\right)
\\
&+
e^{\lambda/2}
\left[
    \frac{
        \ell(\ell+1)e^\nu(\omega_r^2-\omega_i^2)
    }{
        (\omega_r^2+\omega_i^2)^2
    }
    -
    \frac{
        (\epsilon+p)r^2\Gamma_r
    }{
        p(\Gamma_r^2+\Gamma_i^2)
    }
\right]V_r
\\
&+
e^{\lambda/2}
\left[
    \frac{
        \ell(\ell+1)e^\nu\,2\omega_r\omega_i
    }{
        (\omega_r^2+\omega_i^2)^2
    }
    -
    \frac{
        (\epsilon+p)r^2\Gamma_i
    }{
        p(\Gamma_r^2+\Gamma_i^2)
    }
\right]V_i .
\end{split}
\tag{B3}
\end{equation}

\begin{equation}
\label{eq:dUi}
\begin{split}
\frac{dU_i}{dr}
={}&
\frac{(\epsilon+p)g}
     {p(\Gamma_r^2+\Gamma_i^2)}
     \left(-\Gamma_iU_r+\Gamma_rU_i\right)
\\
&+
e^{\lambda/2}
\left[
    -\frac{
        \ell(\ell+1)e^\nu\,2\omega_r\omega_i
    }{
        (\omega_r^2+\omega_i^2)^2
    }
    +
    \frac{
        (\epsilon+p)r^2\Gamma_i
    }{
        p(\Gamma_r^2+\Gamma_i^2)
    }
\right]V_r
\\
&+
e^{\lambda/2}
\left[
    \frac{
        \ell(\ell+1)e^\nu(\omega_r^2-\omega_i^2)
    }{
        (\omega_r^2+\omega_i^2)^2
    }
    -
    \frac{
        (\epsilon+p)r^2\Gamma_r
    }{
        p(\Gamma_r^2+\Gamma_i^2)
    }
\right]V_i .
\end{split}
\tag{B4}
\end{equation}

\begin{equation}
\label{eq:dVr}
\begin{split}
\frac{dV_r}{dr}
={}&
\frac{e^{\lambda/2-\nu}}{r^2}
\left[
    \left(\omega_r^2-\omega_i^2-N_r\right)U_r
    -
    \left(2\omega_r\omega_i-N_i\right)U_i
\right]
\\
&+
\frac{(\epsilon+p)g}{p}
\left[
    \left(
        \frac{1}{\Gamma_{\rm eq}}
        -
        \frac{\Gamma_r}{\Gamma_r^2+\Gamma_i^2}
    \right)V_r
    -
    \frac{\Gamma_i}{\Gamma_r^2+\Gamma_i^2}V_i
\right] .
\end{split}
\tag{B5}
\end{equation}

\begin{equation}
\label{eq:dVi}
\begin{split}
\frac{dV_i}{dr}
={}&
\frac{e^{\lambda/2-\nu}}{r^2}
\left[
    \left(2\omega_r\omega_i-N_i\right)U_r
    +
    \left(\omega_r^2-\omega_i^2-N_r\right)U_i
\right]
\\
&+
\frac{(\epsilon+p)g}{p}
\left[
    \frac{\Gamma_i}{\Gamma_r^2+\Gamma_i^2}V_r
    +
    \left(
        \frac{1}{\Gamma_{\rm eq}}
        -
        \frac{\Gamma_r}{\Gamma_r^2+\Gamma_i^2}
    \right)V_i
\right] .
\end{split}
\tag{B6}
\end{equation}

The complex equations are integrated outward from a small radius near
the stellar center. Regularity gives
\begin{equation}
\label{eq:Ur_center}
    U_r(r\rightarrow0)
    =
    \frac{
        \ell(\omega_r^2-\omega_i^2)
    }{
        (\omega_r^2+\omega_i^2)^2
    }
    Y_0 r^{\ell+1},
\tag{B7}
\end{equation}

\begin{equation}
\label{eq:Ui_center}
    U_i(r\rightarrow0)
    =
    -\frac{
        2\ell\omega_r\omega_i
    }{
        (\omega_r^2+\omega_i^2)^2
    }
    Y_0 r^{\ell+1},
\tag{B8}
\end{equation}

\begin{equation}
\label{eq:Vr_center}
    V_r(r\rightarrow0)
    =
    \frac{\epsilon}{\epsilon+p}
    Y_0 r^\ell ,
\tag{B9}
\end{equation}

\begin{equation}
\label{eq:Vi_center}
    V_i(r\rightarrow0)=0 .
\tag{B10}
\end{equation}

Where we have chosen $Y_0 = 1. $At the stellar surface, the vanishing of the complex Lagrangian
pressure perturbation produces two real boundary conditions:
\begin{equation}
\label{eq:surface_real}
    0
    =
    V_r(R)
    -
    \left.
    \frac{U_r(R)}{2R^2}
    e^{-\lambda(R)/2}
    \frac{d\nu}{dr}
    \right|_{r=R},
\tag{B11}
\end{equation}

\begin{equation}
\label{eq:surface_imag}
    0
    =
    V_i(R)
    -
    \left.
    \frac{U_i(R)}{2R^2}
    e^{-\lambda(R)/2}
    \frac{d\nu}{dr}
    \right|_{r=R}.
\tag{B12}
\end{equation}

For each trial pair \((\omega_r,\omega_i)\), we recompute
\(\Gamma_r(r;\omega)\), \(\Gamma_i(r;\omega)\),
\(N_r(r;\omega)\), and \(N_i(r;\omega)\), and integrate
Eqs.~\eqref{eq:dUr}--\eqref{eq:dVi}. The complex eigenfrequency is
obtained by simultaneously requiring the two surface residuals in
Eqs.~\eqref{eq:surface_real} and \eqref{eq:surface_imag} to vanish.

\section{Relation between the sound-speed difference and the composition gradient}
\label{app:sound_speed_difference}

In the main text we derived the frozen-composition adiabatic sound speed,
\begin{equation}
    c_s^2
    =
    \frac{n_B}{\epsilon+P}
    \left(\frac{\partial P}{\partial n_B}\right)_{s,x_p}.
\end{equation}
Here the derivative is taken at fixed entropy per baryon $s$ and fixed proton
fraction $x_p$. The equilibrium sound speed is instead obtained by taking the
derivative along the beta-equilibrium background, where the proton fraction is
a function of density,
\begin{equation}
    x_p=x_p^\beta(n_B).
\end{equation}
For the isentropic stellar profiles considered in this work, this gives
\begin{equation}
    c_e^2
    =
    \left(\frac{dP}{d\epsilon}\right)_{\beta,s}
    =
    \frac{n_B}{\epsilon+P}
    \left(\frac{dP}{dn_B}\right)_{\beta,s}.
\end{equation}
The total derivative along the beta-equilibrium background can be written as
\begin{equation}
    \left(\frac{d}{dn_B}\right)_{\beta,s}
    =
    \left(\frac{\partial}{\partial n_B}\right)_{s,x_p}
    +
    \frac{dx_p^\beta}{dn_B}
    \left(\frac{\partial}{\partial x_p}\right)_{n_B,s}.
\end{equation}
Therefore,
\begin{equation}
    \left(\frac{dP}{dn_B}\right)_{\beta,s}
    =
    \left(\frac{\partial P}{\partial n_B}\right)_{s,x_p}
    +
    \left(\frac{\partial P}{\partial x_p}\right)_{n_B,s}
    \frac{dx_p^\beta}{dn_B}.
\end{equation}
Substituting this expression into the definition of $c_e^2$ and subtracting
from $c_s^2$ gives
\begin{equation}
\label{eq:cs_ce_difference}
    c_s^2-c_e^2
    =
    -\frac{n_B}{\epsilon+P}
    \left(\frac{\partial P}{\partial x_p}\right)_{n_B,s}
    \frac{dx_p^\beta}{dn_B}.
\end{equation}
This equation shows explicitly that the sound-speed difference is sourced by
the beta-equilibrium composition gradient $dx_p^\beta/dn_B$. If the proton
fraction becomes nearly independent of density, then $dx_p^\beta/dn_B$ becomes
small and the buoyant restoring force is suppressed.

The thermodynamic derivative appearing in Eq.~\eqref{eq:cs_ce_difference} can
be expanded using the Gibbs-Duhem relation,
\begin{equation}
    dP = n_B s\,dT+\sum_i n_i d\mu_i ,
\end{equation}
where $n_i=n_B x_i$. At fixed $n_B$ and $s$, this gives
\begin{equation}
    \left(\frac{\partial P}{\partial x_p}\right)_{n_B,s}
    =
    n_B s
    \left(\frac{\partial T}{\partial x_p}\right)_{n_B,s}
    +
    \sum_i n_B x_i
    \left(\frac{\partial \mu_i}{\partial x_p}\right)_{n_B,s}.
\end{equation}
Hence the sound-speed difference can also be written as
\begin{equation}
\label{eq:cs_ce_difference_expanded}
    c_s^2-c_e^2
    =
    -\frac{n_B^2}{\epsilon+P}
    \left[
    s
    \left(\frac{\partial T}{\partial x_p}\right)_{n_B,s}
    +
    \sum_i x_i
    \left(\frac{\partial \mu_i}{\partial x_p}\right)_{n_B,s}
    \right]
    \frac{dx_p^\beta}{dn_B}.
\end{equation}
For charge-neutral $npe$ matter, the fractions satisfy
\begin{equation}
    x_n=1-x_p, \qquad x_e=x_p .
\end{equation}
Equations~\eqref{eq:cs_ce_difference} and
\eqref{eq:cs_ce_difference_expanded} make clear how the behavior of the
sound-speed difference is connected to the profiles shown in
Fig.~\ref{fig:comp_profiles} and Fig.~\ref{fig:temp_profiles}. The proton-fraction profile enters directly through $dx_p^\beta/dn_B$, while the temperature enters through the thermodynamic derivative $(\partial T/\partial x_p)_{n_B,s}$ and through its effect on the chemical potentials. The density derivative of the temperature profile does not appear as a separate source term in $c_s^2-c_e^2$; the corresponding density derivatives are present in both $c_s^2$ and $c_e^2$ and cancel in the difference.


\bibliographystyle{aasjournal}
\bibliography{main}

\end{document}